\documentclass[bibyear]{aa}

\usepackage{graphicx}
\usepackage{txfonts}
\usepackage{ulem}

\usepackage{hyperref}
\begin{document}

   \title{The large-scale magnetic field of the M~dwarf double-line spectroscopic binary FK~Aqr\thanks{Based on observations obtained at the Canada-France-Hawaii Telescope (CFHT), which is operated by the National Research Council (NRC) of Canada, the Institut National des Sciences de l'Univers of the Centre National de la Recherche Scientifique (CNRS) of France, and the University of Hawaii. The observations at the CFHT were performed with care and respect from the summit of Maunakea, which is a significant cultural and historic site.}}


   \author{S. Tsvetkova\inst{1,2}, J. Morin\inst{1}, C.P. Folsom\inst{3}, J.-B. Le Bouquin\inst{4}, E. Alecian\inst{4}, S. Bellotti\inst{5,6}, G. Hussain\inst{6}, O. Kochukhov\inst{7}, S.C. Marsden\inst{8}, C. Neiner\inst{9}, P. Petit\inst{5}, G.A. Wade\inst{10} \and the BinaMIcS collaboration}

   \institute{LUPM-UMR 5299, CNRS \& Universit\'e Montpellier, Place Eug\`ene Bataillon, 34095 Montpellier, France
   \and Institute of Astronomy and NAO, Bulgarian Academy of Sciences, 72 Tsarigradsko shose, 1784 Sofia, Bulgaria
   \and Tartu Observatory, University of Tartu, Observatooriumi 1, Toravere, 61602, Tartumaa, Estonia
   \and Univ. Grenoble Alpes, CNRS, IPAG, 38000 Grenoble, France
   \and Universit\'e de Toulouse, UPS-OMP, Institut de Recherche en Astrophysique et Plan\'etologie, CNRS, CNES, 14 avenue Edouard Belin, 31400, Toulouse, France
   \and Science Division, Directorate of Science, European Space Research and Technology Centre (ESA/ESTEC), Kepleraan 1, 2201 AZ, Noordwijk, The Netherlands
   \and Department of Physics and Astronomy, Uppsala University, Box 516, 75120, Uppsala, Sweden
   \and University of Southern Queensland, Centre for Astrophysics, West Street, Toowoomba, QLD 4350 Australia
   \and LESIA, Observatoire de Paris, PSL University, CNRS, Sorbonne Universit\'e, Universit\'e Paris Cit\'e, 5 place Jules Janssen, 92195 Meudon, France
   \and Department of Physics, Royal Military College of Canada, PO Box 17000, Station `Forces', Kingston, Ontario, Canada K7K 4B4
   }

   \date{xx; xx}
 
  \abstract
   {This work is part of the BinaMIcS project, the aim of which is to understand the interaction between binarity and magnetism in close binary systems. All the studied spectroscopic binaries targeted by the BinaMIcS project encompass hot massive and intermediate-mass stars on the main sequence, as well as cool stars over a wide range of evolutionary stages.}
   {The present paper focuses on the binary system FK~Aqr, which is composed of two early M~dwarfs. Both stars are already known to be  magnetically active based on their light curves and detected flare activity. In addition, the two components have large convective envelopes with masses just above the fully convective limit, making the system an ideal target for  studying effect of binarity on stellar dynamos.}
   {We use spectropolarimetric observations obtained with ESPaDOnS at CFHT in September 2014. Mean Stokes $I$ and $V$ line profiles are extracted using the least-squares deconvolution (LSD) method. The radial velocities of the two components are measured from the LSD Stokes~$I$ profiles and are combined with interferometric measurements in order to constrain the orbital parameters of the system. The longitudinal magnetic fields $B_{l}$ and chromospheric activity indicators are measured from the LSD mean line profiles. The rotational modulation of the Stokes~$V$ profiles is used to reconstruct the surface magnetic field structures of both stars via the Zeeman Doppler imaging (ZDI) inversion technique.}
   {Maps of the surface magnetic field structures of both components of FK~Aqr are presented for the first time. Our study shows that both components host similar large-scale magnetic fields of moderate intensity ($B_{\rm mean} \simeq 0.25$~kG); both are predominantly poloidal and feature a strong axisymmetric dipolar component.}
   {Both components of FK~Aqr feature a rather strong large-scale magnetic field (compared to single early M~dwarfs with similar masses) with a mainly dipolar axisymmetric structure. This type of magnetic field is not typical for single early M~dwarfs, and is rather reminiscent of fully convective dwarfs with later spectral types. The primary FK~Aqr~A is  currently the most massive recognised main sequence M~dwarf known to host this type of strong dipolar field.}


   \authorrunning{Tsvetkova et al.}
   \titlerunning{Magnetic field topology of FK~Aqr}
   \maketitle
%

\section{Introduction}

Magnetic fields play an important role in  stellar evolution and are found throughout the entire Hertzsprung-Russell (HR) diagram (see \cite{Donati2009} for a review). In particular, cool stars are thought to generate their magnetic field through dynamo action acting within their outer convective envelope, leading to a complex interplay between magnetic field generation, mass loss, and rotational evolution (see e.g. \cite{Brun2017}). Among cool stars, M~dwarfs are main sequence stars covering the mass range between 0.08 and 0.6~M$_{\odot}$, and are the most frequent type of stars in the solar neighbourhood (\cite{Lepine2011}). These stars are of particular interest regarding the generation of their magnetic fields, which involves a transition in their internal structure that takes place during the M~dwarf regime. M~dwarfs earlier than spectral types M3 and M4 or more massive than $\sim 0.35~M_{\odot}$ have a structure similar to that of Sun-like stars ---an inner radiative zone and an outer convective envelope separated by a tachocline. M~dwarfs of later spectral types are fully convective (Chabrier \& Baraffe 1997, \cite{Feiden2021}), and therefore the generation of their magnetic fields may rely on non-solar dynamo processes (see e.g. \cite{Morin2012}). Observational studies show not only that M~dwarfs harbour the strongest magnetic fields among single cool main sequence stars (Shulyak et al. 2017, \cite{Reiners2022}), but also that the topology of their surface magnetic fields change rapidly when crossing the fully convective boundary. Stars more massive than $\sim 0.5~M_{\odot}$ harbour complex multipolar and time-dependant large-scale fields, while those below this threshold host strong dipole-dominated large-scale fields displaying only minor evolution over several years  (Donati et al. 2008, Morin et al. 2008, 2010).

Studying binary and multiple stellar systems gives us a better understanding of stellar magnetic fields. In general, it is assumed that the components of close systems were formed under the same conditions (at the same time and from the same interstellar parental cloud with the same chemical composition). This makes them well-suited targets for investigating the origin, properties, and impact of stellar magnetic fields. The present study is part of the BinaMIcS project (Binary and Magnetic Interactions in various classes of Stars; Alecian et al. 2015~a, b). The aim of BinaMIcS is to study stellar magnetism under the influence of the physical processes occurring in close binary systems. The components of the selected systems cover a large part of the HR diagram, including a wide range of masses on the main sequence from cool low-mass fully convective stars to hot and massive stars. Two key aspects of the BinaMIcS project are that it targets both hot and cool stars in order  (i) to analyse how potential star--star interactions may modify stellar activity with respect to single stars; and (ii) to assess the effect of tidal interactions in close binary systems on the dynamo-generated magnetic fields. Three cool binary systems have already been analysed in the framework of the BinaMIcS project:  $\sigma^2$~CrB (F9V+G0V, \cite{Rosen2018}) V1878~Ori (K2-3+K2-3 weak-line T Tauri stars, \cite{Lavail2020}), and UV~Psc (G5V+K3V, \cite{Hahlin2021}).

The present study focuses on the binary system FK~Aqr (GJ~867~AC, HD~214479), discovered by Vyssotsky \& Mateer (1952) and later studied in detail by Herbig \& Moorhead (1965, HM65 hereafter). FK~Aqr is a double-lined spectroscopic binary with an orbital period of 4.08322~d (HM65) consisting of two M1-2Ve dwarfs of similar mass; their mass ratio is $q=0.8$ (HM65). FK~Aqr is a bright (V=9.09, HM65) and nearby ($\pi=115.01\pm1.30$~mas, Davison et al. 2014) spectroscopic binary, and is part of the quadruple system GJ~867. This is one of only four quadruple systems within 10~pc of the Sun and the only one among these systems with all four cool components (three M~dwarfs and one probable low-mass M~dwarf or brown dwarf; Davison et al. 2014). Indeed, the recent study by Winters et al. (2019) shows that while binaries are fairly common in the solar neighbourhood ---indeed $26.8 \pm 1.4~\%$ of M~dwarf primaries have one or more companions in a volume-limited sample within 25~pc--- quadruple and higher-order systems are much rarer, and make up only $0.3~\%$. FK~Aqr is the primary component of a widely separated visual binary. Its visual companion is FL~Aqr (GJ~867~BD), a single-lined spectroscopic binary of spectral type M3.5Ve. With a visual magnitude of V = 11.45 and a separation of 24”.5, it does not contaminate our observations of FK~Aqr. Taking into account the very similar parallaxes, proper motions, and radial velocities of these two binaries, Davison et al. (2014) consider them to be very likely physically associated.

Both systems show flare activity (Byrne 1978, Byrne 1979, Byrne \& McFarland 1980, Pollock et al. 1991). Pollock et al. (1991) reported that both components contribute to the observed X-ray emission with an intensity ratio A:B of about 3:1. Based on these and other studies, these latter authors concluded that GJ~867~BD flares more frequently but typically produces less-energetic flares than GJ~867~AC in the optical domain (one flare every 30 minutes with integrated energies typically between $10^{29}$ and $5 \times 10^{30}$ erg for GJ~867~BD, and one flare every 4 hours with energies between $10^{31}$ and $10^{32}$~erg for GJ~867~AC). The light curve of FK~Aqr does not exhibit a large-amplitude modulation. The first measurement reported in the literature is presented by Bopp \& Espenak (1977), where $\Delta$V = 0.06 mag. In subsequent studies, the photometric variability is reported to be $\Delta$V < 0.04~mag (Byrne \& McFarland 1980, Byrne et al. 1990). Cutispoto (1995),  Cutispoto \& Leto (1997), and Cutispoto et al. (2003) even report a change in the shape of the light curves from their studies at different epochs, which they interpret as a change in the surface distribution of the spots.

The present paper is organised as follows: A short description of the telescope and observational dataset is given in Section 2. Section 3 describes all the employed methods to process the data and characterise the system: radial velocities of both components are measured and combined with astrometric measurements in order to fully characterise the orbit of the system; longitudinal magnetic field values $B_{l}$ are computed from mean Stokes~$I$ and $V$ LSD profiles; and the three classical chromospheric activity indicators (H${\alpha}$, CaII H\&K, and CaII IRT) are measured. In Section 4 we describe how the surface magnetic field topology of both stars is modelled from the Stokes~$V$ mean LSD profiles with the Zeeman Doppler imaging (ZDI) method. We discuss our results and provide concluding remarks in Section 5.

\section{Observations}

\subsection{ESPaDOnS spectro-polarimetry}
Observational data were obtained with the fibre-fed \'echelle spectropolarimeter ESPaDOnS (Donati et al. 2006), which operates at the 3.6m Canada-France-Hawaii Telescope (CFHT). In polarimetric mode, the instrument has a spectral resolving power of about 65~000 and an almost continuous spectrum coverage from the near-ultraviolet (at about 370 nm) to the near-infrared domain (at about 1050~nm) in a single exposure, with 40 orders aligned on the CCD frame by two cross-disperser prisms. The Stokes~$I$ (unpolarised light) and Stokes~$V$ (circular polarisation) parameters are simultaneously measured using a sequence of four subexposures, between which the retarders (Fresnel rhombs) are rotated in order to exchange the beams in the instrument and to reduce spurious polarisation signatures (Semel et al. 1993).

The automatic reduction software \textsc{Libre-ESpRIT} (Donati et al. 1997) is applied to all observations as a component of the CFHT Upena reduction pipeline, which includes the optimal extraction of the spectrum, wavelength calibration, correction to the heliocentric frame, and continuum normalisation.

In total, 26 polarimetric sequences were obtained for FK~Aqr under the BinaMIcS project, covering the observational period 3--16 September 2014. The total exposure time per spectrum is 960~s (4 x 240~s). The observations reach a peak signal-to-noise ratio (S/N) in Stokes~$I$ between 233 and 731 at 871~nm. The journal of observations is presented in Table~\ref{table:journal}, where column 1 gives the date of observation, column 2 gives the heliocentric Julian date of the corresponding observation, column 3 gives the rotation cycles, column 4 gives the rotational phases, calculated assuming a rotational period of 4.08319598 days (this is a refined value of the period, which we obtained with the \textsc{Phoebe} code, which is described in detail in Sect.~\ref{subsec:phoebe}). The remaining columns give the measurements of the longitudinal magnetic field and the radial velocity of both components of FK~Aqr, which are described in the following section. 

\begin{table*}

\centering
\begin{center}

\caption{Journal of observations and measurements of the longitudinal magnetic fields $B_{l}$ and radial velocities of both components of the close binary FK~Aqr. The symbols `p' and `s' indicate the primary and secondary components, respectively.}

\label{table:journal}

\centering
\begin{tabular}{c c c c c c c c c c c c}
\hline\hline

Date  &    HJD       & Rot.  & Rot.  & $B_{l}(p)$ & $\sigma$ & $B_{l}(s)$ & $\sigma$ & $RV(p)$          & $\sigma$         & $RV(s)$          & $\sigma$\\
UT    &2\,456\,900 + & cycle & phase &  [G]       &   [G]    &   [G]      &  [G]     & [km\,s$^{-1}$]   & [km\,s$^{-1}$]   & [km\,s$^{-1}$]   & [km\,s$^{-1}$]\\
\hline
(1)   & (2)          & (3)  & (4)        & (5)      & (6)        &     (7)  & (8)      &  (9)     &  (10)    & (11) & (12)\\
\hline
\\
03 Sep 2014 & 03.86539 & 4839 & 0.0307 &         &       &         &      & -16.70 & 0.06 & 3.40   & 0.11\\
04 Sep 2014 & 04.03408 & 4839 & 0.0720 & -66.5   & 6.0   & -266.2  & 19.8 & -28.14 & 0.04 & 18.58  & 0.11\\
04 Sep 2014 & 04.79419 & 4839 & 0.2581 & -108.3  & 6.0   & -81.8   & 21.9 & -53.52 & 0.04 & 50.86  & 0.10\\
04 Sep 2014 & 04.89452 & 4839 & 0.2827 & -124.4  & 5.7   & -85.0   & 21.1 & -52.53 & 0.04 & 49.69  & 0.10\\
05 Sep 2014 & 05.92014 & 4839 & 0.5339 &         &       &         &      & 2.53   & 0.07 & -19.39 & 0.13\\
06 Sep 2014 & 06.03950 & 4839 & 0.5631 &         &       &         &      & 10.78  & 0.05 & -29.87 & 0.22\\
06 Sep 2014 & 06.93620 & 4839 & 0.7827 & -126.4  & 6.1   & -315.5  & 24.1 & 37.61  & 0.04 & -64.24 & 0.08\\
07 Sep 2014 & 07.03896 & 4839 & 0.8079 & -136.6  & 6.0   & -367.7  & 24.8 & 35.50  & 0.04 & -61.22 & 0.09\\
07 Sep 2014 & 07.83861 & 4840 & 0.0037 &         &       &         &      & -8.28  & 0.04 &        &     \\
07 Sep 2014 & 07.98344 & 4840 & 0.0392 &         &       &         &      & -19.17 & 0.05 & 6.66   & 0.14\\
08 Sep 2014 & 08.81651 & 4840 & 0.2432 & -89.7   & 8.8   & -87.5   & 31.7 & -53.63 & 0.04 & 50.90  & 0.11\\
08 Sep 2014 & 08.83144 & 4840 & 0.2469 & -78.7   & 10.0  & -161.3  & 37.5 & -53.64 & 0.04 & 50.92  & 0.11\\
08 Sep 2014 & 08.96914 & 4840 & 0.2806 & -105.7  & 7.1   & -27.4   & 26.3 & -52.65 & 0.04 & 49.85  & 0.10\\
08 Sep 2014 & 08.98421 & 4840 & 0.2843 & -116.5  & 7.7   & -35.7   & 28.6 & -52.40 & 0.04 & 49.58  & 0.10\\
09 Sep 2014 & 09.80282 & 4840 & 0.4848 &         &       &         &      & -9.73  & 0.09 &        &     \\
09 Sep 2014 & 09.94937 & 4840 & 0.5207 &         &       &         &      & -1.70  & 0.11 &        &     \\
10 Sep 2014 & 10.80926 & 4840 & 0.7313 & -142.8  & 16.5  & -291.7  & 58.0 & 38.39  & 0.04 & -65.40 & 0.09\\
10 Sep 2014 & 10.82560 & 4840 & 0.7353 & -135.0  & 9.7   & -323.7  & 36.2 & 38.47  & 0.04 & -65.55 & 0.09\\
10 Sep 2014 & 10.84093 & 4840 & 0.7390 & -117.6  & 18.4  & -325.5  & 64.3 & 38.56  & 0.05 & -65.63 & 0.09\\
10 Sep 2014 & 10.92491 & 4840 & 0.7596 & -110.2  & 5.8   & -329.9  & 24.4 & 38.52  & 0.04 & -65.55 & 0.08\\
11 Sep 2014 & 11.81364 & 4840 & 0.9772 &         &       &         &      & -2.17  & 0.11 &        &     \\
11 Sep 2014 & 11.94741 & 4841 & 0.0100 &         &       &         &      & -9.52  & 0.07 &        &     \\
11 Sep 2014 & 11.96231 & 4841 & 0.0137 &         &       &         &      & -10.49 & 0.12 &        &     \\
12 Sep 2014 & 12.81128 & 4841 & 0.2216 & -60.4   & 6.5   & -120.5  & 25.3 & -53.04 & 0.04 & 50.21  & 0.10\\
12 Sep 2014 & 12.97476 & 4841 & 0.2616 & -87.6   & 6.5   & -53.5   & 24.2 & -53.51 & 0.04 & 50.72  & 0.10\\
16 Sep 2014 & 16.99923 & 4842 & 0.2472 & -66.9   & 6.2   & -45.0   & 22.4 & -53.67 & 0.04 & 50.96  & 0.10\\

\\
\hline
\hline
\end{tabular}
\end{center}
\end{table*}

\subsection{PIONIER near-infrared interferometry}
\label{subsec:pionier}

Optical interferometric observations were obtained with the PIONIER near-infrared instrument (\cite{LeBouquin2011}) of the Very Large Telescope Interferometer (VLTI). These observations allows us to spatially resolve the components of the binary at 16 different orbital positions during the period of time from 9 June 2014 to 2 September 2014. The data have been reduced by the standard \texttt{pndrs} pipeline of the instrument described by \cite{LeBouquin2011}.

Each observation has been adjusted with a simple binary model made of two uniform discs. The disc diameters have fixed values of 0.58\,mas and 0.48\,mas based on the expected apparent diameters of M2 dwarfs at a distance of 8.89\,pc (from the Gaia parallax). These diameters are only marginally resolved by the baselines of VLTI, and therefore their exact values have negligible impact on the estimated separation and flux ratio.

The list of best-fit flux ratio, separation, and position angle for each interferometric observation is summarised in Table~\ref{table:journal_astrometry}. The typical uncertainty on the separation is $0.1\,$mas, and therefore  the  separation of a few milliarcseconds between the two components is very well resolved. On average, the flux ratio between the two components in the near-infrared H-band is $f_2/f_1 = 0.47\pm0.01$.

\begin{table*}
\centering
\begin{center}
\caption{Journal of observations and measurements of the interferometric observations. The flux ratio is given in the H-band ($1.4-1.7\,\mu$m).}

\label{table:journal_astrometry}

\begin{tabular}{c c c c c c c}
\hline\hline
HJD      & flux ratio & separation & position angle & $e_{min}$ & $e_{max}$ & PA $e_{max}$\\
2\,400\,000.5 +& & [mas] & [deg] & [mas] & [mas] & [deg] \\
\hline
\\
 56868.297 &  0.482 &   3.78 &  +170.54 &    0.10 &   0.07 &     140 \\
 56869.285 &  0.501 &   5.37 &   -83.29 &    0.10 &   0.05 &     101 \\
 56869.302 &  0.479 &   5.40 &   -82.42 &    0.09 &   0.04 &     111 \\
 56869.385 &  0.492 &   5.53 &   -77.72 &    0.12 &   0.07 &     135 \\
 56870.330 &  0.473 &   3.77 &   -10.75 &    0.07 &   0.03 &      12 \\
 56870.344 &  0.471 &   3.74 &    -8.99 &    0.05 &   0.03 &      10 \\
 56871.315 &  0.486 &   5.35 &   +95.83 &    0.16 &   0.06 &     129 \\
 56871.329 &  0.484 &   5.37 &   +96.88 &    0.16 &   0.07 &     135 \\
 56871.425 &  0.481 &   5.55 &  +102.29 &    0.33 &   0.14 &     125 \\
 56872.317 &  0.473 &   3.95 &  +162.88 &    0.09 &   0.05 &     178 \\
 56872.347 &  0.474 &   3.83 &  +166.38 &    0.13 &   0.08 &       9 \\
 56873.253 &  0.493 &   5.11 &   -90.33 &    0.20 &   0.12 &     119 \\
 56873.270 &  0.487 &   5.13 &   -89.68 &    0.17 &   0.08 &     117 \\
 56873.355 &  0.469 &   5.32 &   -84.31 &    0.12 &   0.07 &     131 \\
 56873.371 &  0.472 &   5.34 &   -83.64 &    0.12 &   0.06 &     128 \\
 56903.150 &  0.485 &   3.47 &    +9.65 &    0.12 &   0.07 &     170 \\
\\
\hline
\hline

\end{tabular}
\tablefoot{The position angle is the position of the secondary (faintest in H-band), measured north (0~deg) to east (90~deg). $e_{min}$ and $e_{max}$ are the semi-minor and the semi-major axes of the 1$\sigma$ astrometric error ellipse. PA $e_{max}$ is the position angle of the semi-major axis of the error ellipse, measured from north (0\,deg) to east (90\,deg).}

\end{center}
\end{table*}

\section{Analysis}
\subsection{Least-squares deconvolution method}
\begin{figure*}
    \sidecaption
    \includegraphics[width=0.30\textwidth]{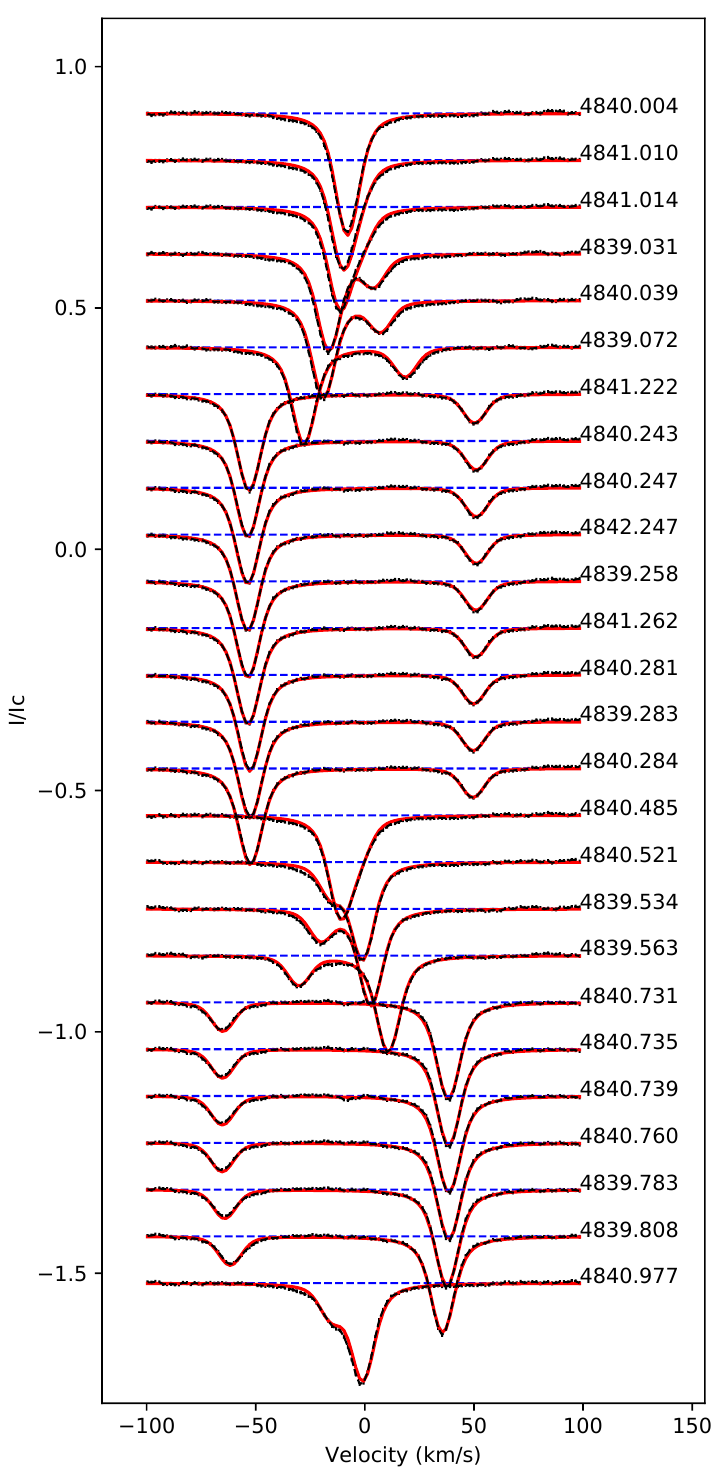}
    \includegraphics[height=0.45\textheight]{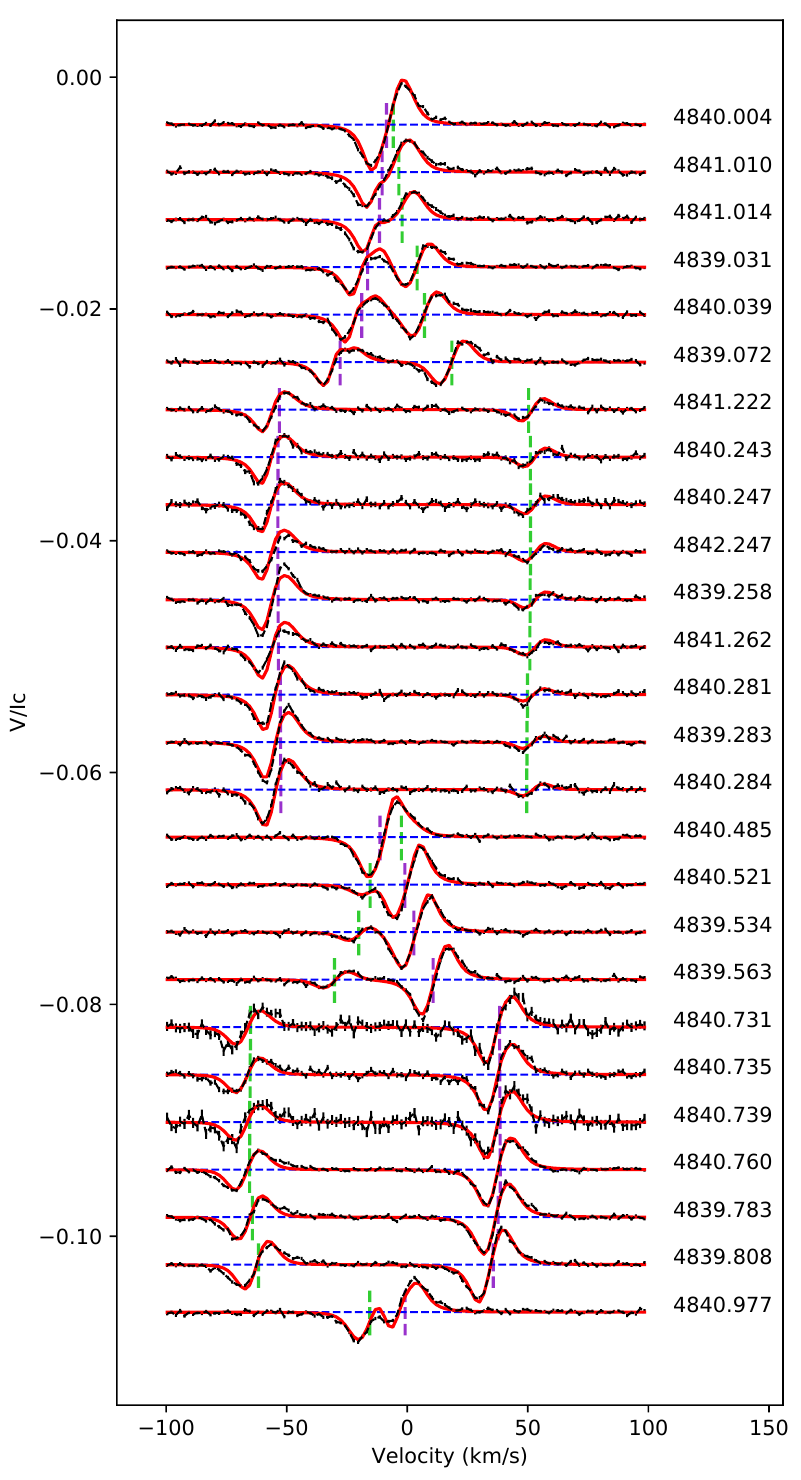}
      \caption{Normalised Stokes~$I$ and $V$ profiles of FK~Aqr in the observational period 3--16 September 2014. Observed profiles are plotted in dashed black lines, synthetic profiles are in red lines, and blue horizontal dashed lines are the zero level. All profiles are shifted vertically for display purposes. The rotational phases of the observations are indicated in the right part of the plot next to each profile. The vertical violet (for the primary) and green (for the secondary) dashed lines in the Stokes $V$ panel indicate the center of the line profiles.}
   \label{fig:FKAqrStokesIV}
   \end{figure*}

We used the least-squares deconvolution (LSD) multi-line technique (Donati et al. 1997, \cite{Kochukhov2010}) to generate the mean Stokes~$I$ and $V$ line profiles\footnote{We use the Python implementation developed by C.P.~Folsom: \url{https://github.com/folsomcp/LSDpy}}. In general, this technique averages several thousand photospheric atomic lines from one spectrum in order to increase the S/N, thereby allowing us to detect weak polarised Zeeman signatures. In the case of FK~Aqr, we applied a line mask that is computed for M~dwarfs and is already used over a sample of this type of star (Morin et al. 2010). The mask is created from an Atlas9 local thermodynamic equilibrium model (Kurucz 1993) with $T_{\rm eff}=3\,500~K$ and $\log g=5$ and a threshold in the line depth of 40~\% (Donati et al. 2008). As a result, around $5\,000$ spectral lines were averaged per spectrum. The LSD profiles were calculated with the following normalisation parameters: normalised line depth equal to 0.645, wavelength equal to 722~nm, and an effective Land\'e factor equal to 1.20. The Stokes~$I$ profiles exhibit a pseudo-continuum level of below unity. This effect is observed in all M~dwarfs and is believed to be caused by the contribution of molecular spectral lines that are not included in the line list; we therefore scaled it in order to bring it to unity. The profiles are presented in Fig.~\ref{fig:FKAqrStokesIV}. Clear Stokes~$V$ signatures are visible and detected from all observations of both components. Their shapes are all the same for both stars: simple two-lobe shapes with negative blue and positive red lobes.

The observations are phased according to the following ephemeris:

\begin{equation}
 HJD = 2437145.1548 + 4.08319598\, \phi
 \label{eq:ephemeris}
,\end{equation}
where HJD is the heliocentric Julian date of the observations and $\phi$ is the rotational cycle. The values of $T_0 = 2437145.1548 \pm 0.0046$ and $P_{orb} = 4.08319598 \pm 0.00000095$~d were obtained as detailed in Sect.~\ref{subsec:phoebe}. As FK~Aqr is a synchronised close binary, the rotational periods of both components are the same. Calculated in this way, rotational phases are listed in column 4 of Table~\ref{table:journal}.

\subsection{Radial velocity}
Radial velocities (RV) of both components of the system are measured from their LSD Stokes~$I$ profiles by fitting a Lorenzian function. The error bars contain both the uncertainty from the fit, and the stability of the ESPaDOnS spectropolarimeter, which is 30~m\,s$^{-1}$ (Moutou et al. 2007). The measured values and their calculated error bars are given in  columns 9 to
12 of Table~\ref{table:journal}. We used these radial velocity measurements combined with the measurements given by HM65 (and the corresponding error bars of both datasets) to refine the orbital parameters of the system employing the \textsc{Phoebe} code. This is described in detail in Sect.~\ref{subsec:phoebe}.

\subsection{Astrometry of the binary FK~Aqr}
\label{subsec:astrometry}

\begin{figure}
    \centering
    \includegraphics[width=\columnwidth]{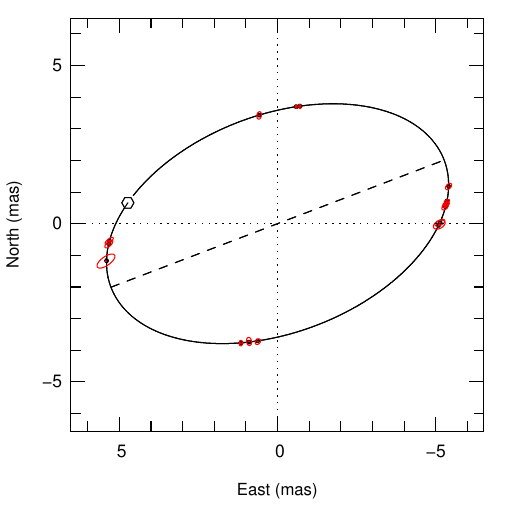}
     \caption{Best-fit orbital solution to the astrometric observations. The orbital trace is the motion of the secondary around the primary (the primary being the brightest in the H-band). The periastron of the secondary is represented by a filled symbol and the line of nodes by a dashed line. The astrometric observations are the red ellipses (of typical size less than 0.2\,mas), and the corresponding points along the orbit are the black dots. North is up and east is left.}
   \label{fig:astrometry_best_fit}
   \end{figure}

We ran an orbital fit of the astrometric positions. There are many tools and methods available to perform such a fit. Here, we used the code already presented in \cite{LeBouquin2017}. The best-fit parameters are summarised in Table~\ref{table:best_fit_astrometry_alone} and the orbital trace is shown in Figure~\ref{fig:astrometry_best_fit}.

We found that the eccentricity $e$ is small and compatible with zero. Consequently, the argument of periastron of the secondary $\omega{}$ is poorly defined. Because we kept this parameter free, the time of periastron passage $T{}$ is also poorly defined, as one can be converted into the other by a rotation of the circular orbit. Nevertheless, this is not a problem for this analysis, where we are mostly interested in the inclination $i=52.39\pm0.45\,$deg and in the semi-major axis of the apparent orbit $a=5.635\pm0.037\,$mas. These two parameters are very well constrained thanks to the dense sampling of the astrometric orbit.

It is possible to compute the total mass of the system with Kepler's law by combining the apparent size of the astrometric orbit, the orbital period, and the Gaia distance ($d=8.897\pm0.004$~pc) (Gaia Collaboration 2020). We found $M_t=1.008\pm0.020\,{M}_\odot$.

\begin{table}
\centering
\caption{Best-fit parameters of the astrometric orbit.}
\label{table:best_fit_astrometry_alone}
 \begin{tabular*}{6.35cm}{cccc}
 \hline\hline\noalign{\smallskip}
 Element & Unit & Value & Uncertainty \\
 \noalign{\smallskip}\hline\noalign{\smallskip}
 $T$ & MJD  &  $37145.1$  &  $2.2$  \\
 $P$ & days  &  $4.08322$  &  $0.00068$  \\
 $a$ & mas  &  $5.635$  &  $0.037$  \\
 $e$ &   &  $0.0000$  &  $0.0029$  \\
 $\Omega$ & deg  &  $110.89$  &  $0.47$  \\
 $\omega$ & deg  &  $318$  &  $121$  \\
 $i$ & deg  &  $52.39$  &  $0.45$  \\
 \noalign{\smallskip}\hline\noalign{\smallskip}
 \multicolumn{4}{c}{From apparent orbit and Gaia distance}\\
 $M_t$ & ${M}_\odot$  &  $1.008$  &  $0.020$  \\
 \noalign{\smallskip}\hline
 \end{tabular*}

\end{table}

We also performed a simultaneous fit of the combined radial velocities (those of HM65 and those of 2014 ESPaDOnS observations) and astrometric observations using the code of \cite{LeBouquin2017}. This is presented in Appendix~\ref{appendix:astrometry_and_RV}.

\subsection{Refining the orbital parameters with \textsc{Phoebe}}
\label{subsec:phoebe}

As the ESPaDOnS spectropolarimeter has excellent stability, which allows very precise measurement of stellar radial velocities, we can use our measurements to refine the orbital solution of the close binary FK~Aqr. There is only one orbital solution published in the literature, that given by HM65. To this end, we used the \textsc{Phoebe}\footnote{\url{http://phoebe-project.org/}} binary modeling code (Pr$\check{s}$a \& Zwitter 2005, Pr$\check{s}$a et al. 2016). \textsc{Phoebe} uses the MIT licensed \textsc{emcee} backend function (Foreman-Mackey et al. 2013), which is a pure-Python implementation of Goodman \& Weare’s Affine Invariant Markov chain Monte Carlo (MCMC) ensemble sampler. We applied this \textsc{emcee} backend function to our 2014 ESPaDOnS dataset alone, to the HM65 dataset alone, and also to the two datasets combined into one.

As a very first inspection of the radial velocity dataset, we applied a Lomb-Scargle period search analysis (Lomb 1976, Scargle 1982) over the combined dataset (HM65 dataset combined with our dataset from 2014 ESPaDOnS). Given the long temporal gap between the two datasets, the periodogram of the combined dataset for the orbital period displays numerous aliases of similar likelihood. Therefore, given the very small uncertainty that \textsc{Phoebe} calculates for the period (below), we consider that the HM65 period is very reliable.


Employing the \textsc{emcee} function of \textsc{Phoebe}, we initially  started with shorter chains to inspect the datasets; we ran the HM65 dataset alone, the 2014 ESPaDOnS dataset alone, and then both combined as one. We found that 30 walkers and $7\,000$ iterations were sufficient to achieve very good values for the parameters of the orbital solution and a reduced $\chi^{2}$ of the fit of between 1.0 and 2.0, both within a reasonable amount of computation time. This allowed us to run more than 25 models. In different models, we gave  the parameters and their intervals of exploration different central values. Moreover, in some models, we fixed the values of certain orbital parameters and allowed \textsc{Phoebe} to work with fewer unknowns. We used Gaussian distributions. In general, \textsc{Phoebe}'s values of the orbital parameters were consistent with the values provided by  HM65, except for the systemic velocity. The outcome of \textsc{Phoebe}'s \textsc{emcee} function is as follows: (1) Applied to the HM65 dataset alone, \textsc{Phoebe}'s orbital solution matches the orbital solution given by HM65  within the error bars, and is consistent with the radial velocity precision estimated by these latter authors of 1.4~km\,s$^{-1}$. (2) Applied to the 2014 ESPaDOnS dataset alone, \textsc{Phoebe}'s orbital solution matches  the orbital solution given by HM65 within the error bars, except for the systemic velocity of $V_\gamma \simeq -7.36 \pm 0.02$~km\,s$^{-1}$, which is different from the HM65 value of -8.7~km\,s$^{-1}$ (no error bar is given in HM65) by approximately 1.3~km\,s$^{-1}$. (3) Applied to a combined dataset, there is again a difference between the two systemic velocities. A change in radial velocity over several decades could be explained by secular acceleration (K\"urster et al. 2003) or orbital evolution of the wide system GJ~867~AC/BD (Davison et al. 2014), but none of these effects alone or combined could produce a variation of 1.3~km\,s$^{-1}$ over a time span of 60~years. We therefore assume that the difference is largely explained by the accuracy of the HM65 radial velocity measurements, which are based on chromospheric emission lines. Different lines (or wavelength intervals) are used in RV measurements, and there are offsets to be considered (due to slightly different line asymmetries). The systematic difference between the RVs from ESPaDOnS and HM65 may also come from this. We therefore shift the RV values of HM65 by 1.4~km\,s$^{-1}$ (the radial velocity precision given by HM65).

\begin{figure*}
    \centering
    \includegraphics[scale=0.4]{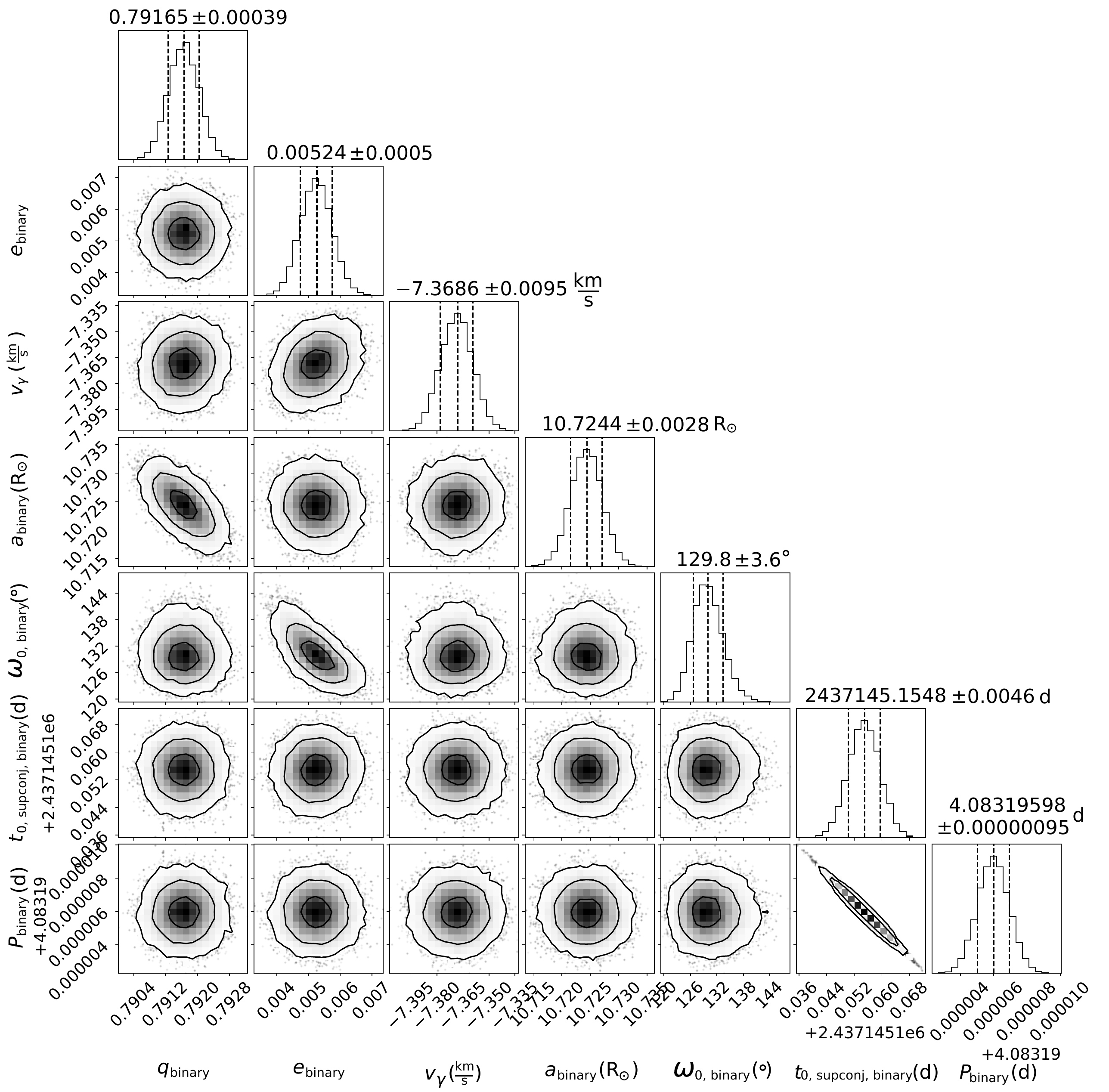}
     \caption{Corner plot of the posterior distributions for the \textsc{Phoebe} RV model of FK~Aqr using the \textsc{emcee} MCMC sampler. The dashed black lines correspond to the 16th, 50th, and 84th percentiles.}
   \label{fig:corner_plot}
   \end{figure*}

\begin{figure}
    \centering
    \includegraphics[scale=0.38]{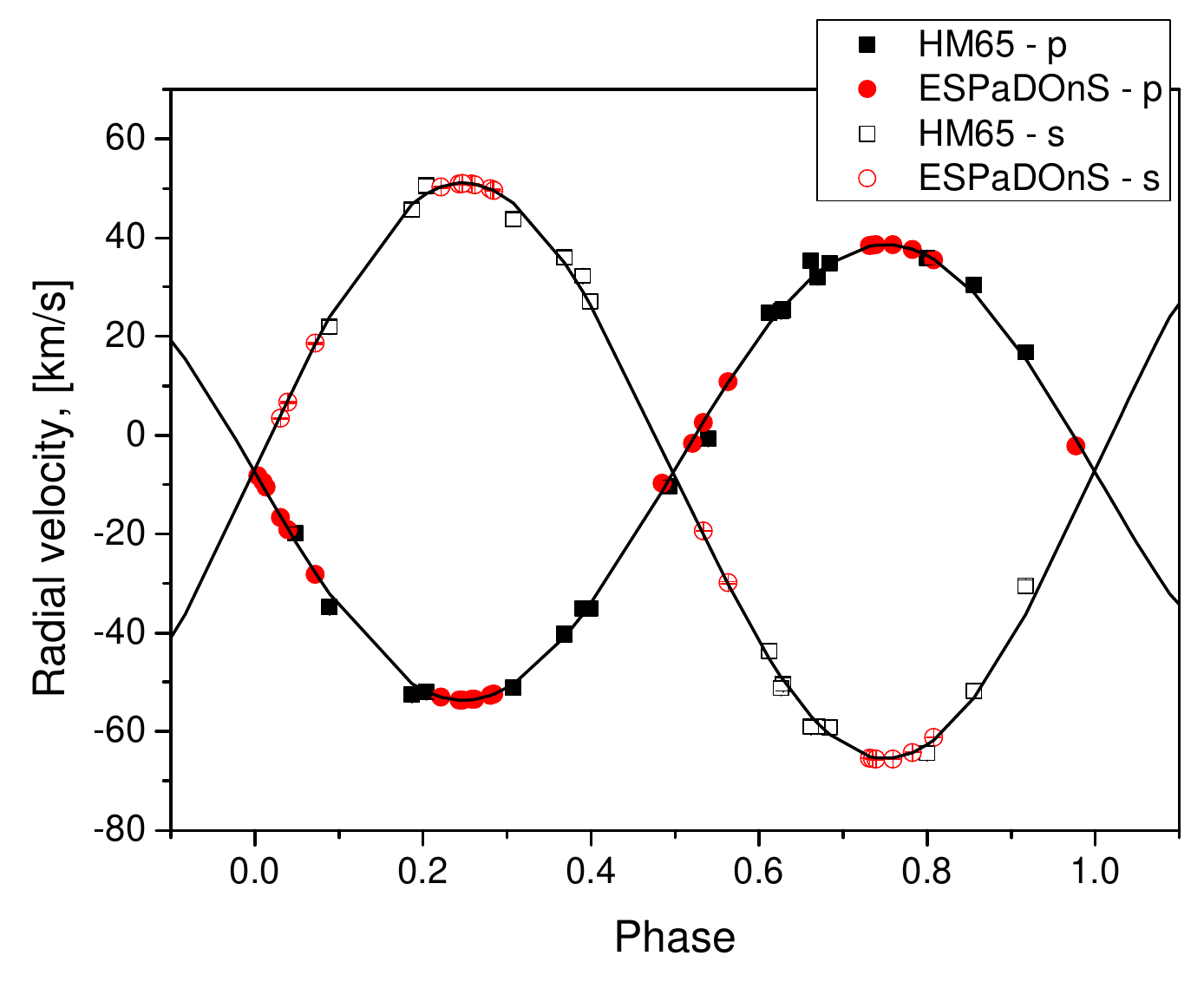}
    \includegraphics[scale=0.38]{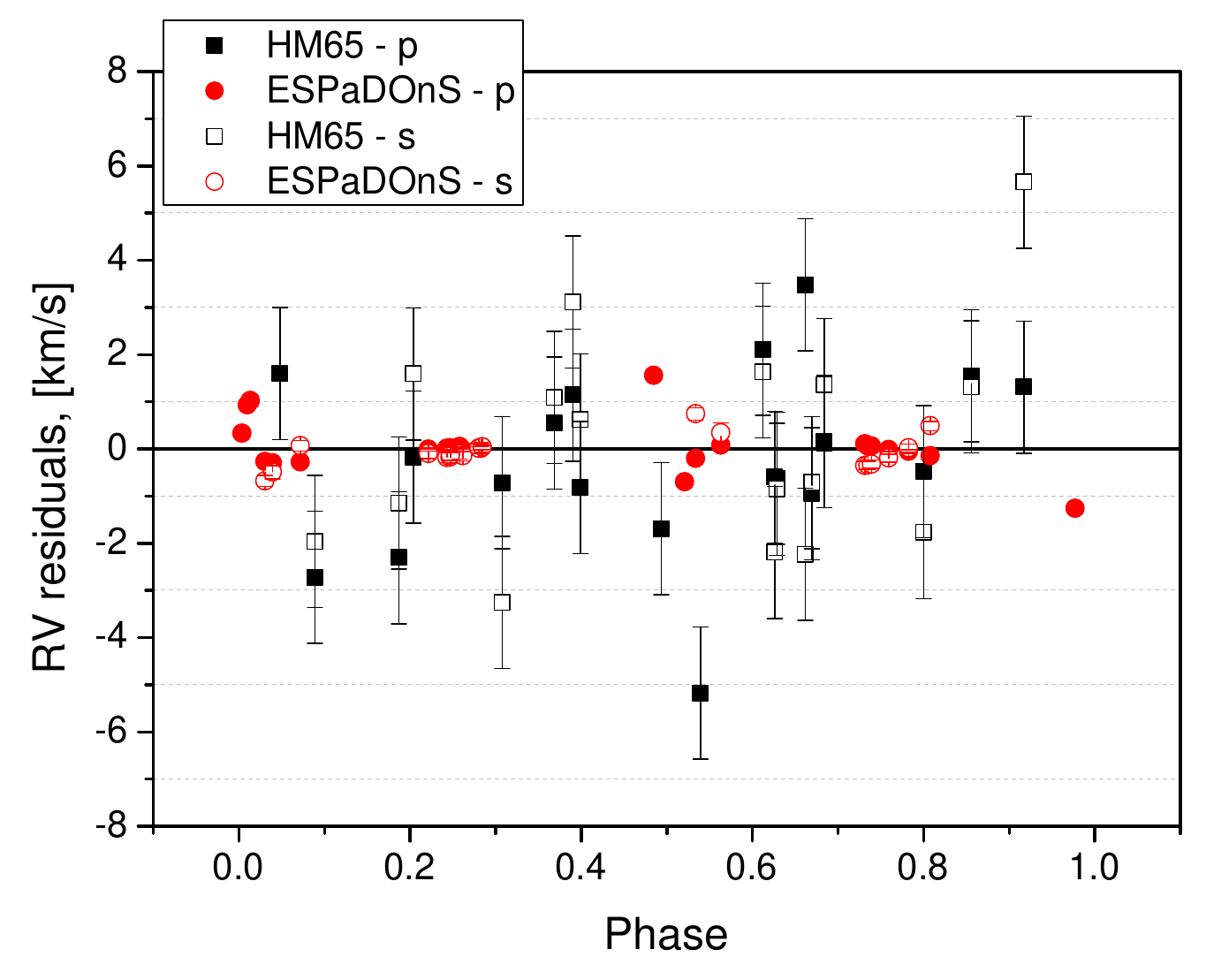}
     \caption{\textsc{Phoebe}'s orbital solution of FK~Aqr based on RV measurements. Upper panel: RV measurements of both components are given with symbols according to the legend in the upper right corner: symbols `p' and `s' stand for the primary and secondary, respectively. The Phoebe model is given in black lines. Lower panel: RV residuals according to the same legend given in the upper left corner. In both panels, all error bars are plotted, but in most cases they are within the symbols of the measurements.}
   \label{fig:RV}
   \end{figure}

To achieve the final results presented in this paper, we ran the \textsc{emcee} function of  \textsc{Phoebe} with 30 walkers and $50\,000$ iterations over the combined dataset. We fixed the value of the angle $i = 52.39^\circ \pm 0.45^\circ$ (derived in Sec.~\ref{subsec:astrometry}). Thus, the parameters over which \textsc{Phoebe} iterated, the priors, are -- $q = 0.8 \pm 0.2$, $e = 0.01 \pm 0.01$, $V_\gamma = -8.0 \pm 2.0$~km\,s$^{-1}$, $a_{binary} = 10.8 \pm 2.0~R_\odot$, $\omega = 356^\circ \pm 60^\circ$, $T_0 = 2437145.143805 \pm 0.005,$ and $P_{orb} = 4.08322 \pm 0.00007$~d. The corner plot of the \textsc{Phoebe} calculations is displayed in Fig.~\ref{fig:corner_plot}. The final results are given in Table~\ref{table:phoebe}. The last column of this table lists the output of \textsc{Phoebe} and the error bars, and the middle column shows the HM65 orbital solution for comparison. The \textsc{Phoebe} model and all the radial velocity measurements are shown in the upper panel of Fig.~\ref{fig:RV}, while the lower panel of the same figure shows the RV residuals. The reduced $\chi^2$ of the fit is 1.33 for the primary and 1.55 for the secondary. The root mean square (rms)  of the residuals of the ESPaDOnS measurements is 0.5 for the primary and 0.3 for the secondary. The rms of the residuals of the HM65 measurements is 1.9 for the primary and 2.3 for the secondary.

As described above, we present \textsc{Phoebe}'s orbital solution based on the combined dataset (archival RVs of HM65 and RVs of 2014 ESPaDOnS dataset). In this combined dataset, we shifted the RV values of HM65 by 1.4~km\,s$^{-1}$ because of the discrepancy we noticed in the systemic velocity. For the completeness of this study, we ran \textsc{Phoebe} again, this time using the original RV values given by HM65 (not corrected by 1.4~km\,s$^{-1}$), and under the same conditions as we used for presenting the final orbital solution: combined dataset, 30 walkers, 50~000 iterations, and the same central values and intervals for the parameters. This did not lead to a change in the values of the orbital parameters, which are listed in Table~\ref{table:phoebe}, but we obtained slightly worse reduced $\chi^2$ values: 1.69 for the primary and 1.78 for the secondary.

\begin{table}
\centering
\begin{center}


\caption{Orbital parameters derived from radial velocity measurements.} 

\label{table:phoebe}

\centering
\begin{tabular}{@{\,}l@{\,\,\,\,}l@{\,\,\,\,}l@{\,}}
\hline\hline

 & Herbig \& Moorhead &  This study \\
 &  (1965)            &   \\
\hline

\\
P$_{orb}$ (days)              & 4.08322 $\pm$ 0.00004      & 4.08319598 $\pm$ 0.00000095 \\
T$_{0}$                       & 2~437~144.123 $\pm$ 0.006  & 2~437~145.1548 $\pm$ 0.0046 \\
e                             & 0.010 $\pm$ 0.010          & 0.0052 $\pm$ 0.0005 \\
$V_\gamma$ (km\,s$^{-1}$)      & -8.7                       & -7.3686 $\pm$ 0.0095 \\
K$_{A}$ (km\,s$^{-1}$)         & 46.8 $\pm$ 0.45            & 46.14 $\pm$ 0.07 \\
K$_{B}$ (km\,s$^{-1}$)        & 58.1 $\pm$ 0.46            & 58.28 $\pm$ 0.10 \\
$\omega$ (deg)                & 356 $\pm$ 32               & 129.8 $\pm$ 3.6 \\
$q=M_{B}/M_{A}$               & 0.80                       & 0.79165 $\pm$ 0.00039 \\
M$_{A}sin^{3}i$ (M$_{\odot})$ & 0.271 $\pm$ 0.006          & 0.267 $\pm$ 0.005 \\
M$_{B}sin^{3}i$ (M$_{\odot})$ & 0.218 $\pm$ 0.005          & 0.214 $\pm$ 0.005 \\
a$_{A}sini$ (km)              & 2.63 ($\pm$ 0.03) x $10^6$ & 2.59 ($\pm$ 0.004) x $10^6$ \\
a$_{B}sini$ (km)              & 3.26 ($\pm$ 0.03) x $10^6$ & 3.27 ($\pm$ 0.006) x $10^6$ \\

\\
\hline

\end{tabular}
\tablefoot{Our values were derived with \textsc{Phoebe} using the combined dataset which includes our 2014 ESPaDOnS dataset and the HM65 dataset. Note that different definitions of the argument of periastron in HM65 and our analysis result in a $180^{\circ}$ difference for the argument of periastron.}

\end{center}
\end{table}

\subsection{$v\sin i$ values}
\label{subsec:vsini_values}

The projected rotational velocities ($v\sin i$ values) of both components of the binary are among the input parameters needed in order to reconstruct their magnetic maps. The catalogue of Eker et al. (2008) provides $v\sin i$ values of 7~km\,s$^{-1}$ for both components. Houdebine (2008) gives a value of 4.7~km\,s$^{-1}$ with an accuracy of 2~km\,s$^{-1}$ using the method of cross-correlation between the target stars and three template stars. Later, Houdebine (2010) measured $v\sin i$  again using observations from HARPS and ELODIE and the corresponding values are 7.17~km\,s$^{-1}$ from HARPS and 4.7 and 7.3~km\,s$^{-1}$ from ELODIE with an uncertainty of 0.3~km\,s$^{-1}$. These latter authors adopt a final value of 7.02~km\,s$^{-1}$.

From the astrometric value of the inclination angle~$i$ and the orbital solution achieved with the \textsc{Phoebe} code, we can easily calculate the masses of the two stars -- $0.55 \pm 0.01$~M$_{\odot}$ and $0.44 \pm 0.01$~M$_{\odot}$ for the primary and secondary, respectively. We then interpolated the values of the masses and radii of the M~dwarfs given by Pecaut \& Mamajek (2013) to find the radii of the two components of our system. We found values equal to $R_A=0.56~{\rm R_{\odot}}$ and $R_B=0.45~{\rm R_{\odot}}$ for the primary and secondary, respectively. From the values of stellar radius, rotational period, and inclination angle, we calculated $v\sin i$ values for both components, finding $5.4 \pm 0.6$~km\,s$^{-1}$ and $4.4 \pm 0.6$~km\,s$^{-1}$ for the primary and secondary, respectively. We use these two values in our modelling of the magnetic field topologies of the two stars as described in the following section.

\subsection{Mean longitudinal magnetic field $B_{l}$}

We computed the line-of-sight component of the stellar magnetic field integrated over the visible stellar disc, $B_{l}$, using the first-order moment method (Rees \& Semel 1979, Donati et al. 1997,  Wade et al. 2000~a,b) and the following equation:
\begin{equation}
 B_{l} =-2.14\times10^{11}\frac{\int vV(v)\,dv}{\lambda gc \int [1-I(v)]\,dv}
,\end{equation}
where $v$ (km\,s$^{-1}$) is the radial velocity in the stellar restframe, $\lambda$ (in nm) is the normalisation wavelength (722~nm for FK~Aqr), $g$ is the Land\'e factor (here 1.20), and c (km\,s$^{-1}$) is the light velocity in vacuum. We set the velocity boundaries of the integration window at 56~km\,s$^{-1}$ around the line centre of the LSD profiles for both components of the binary. We compute $B_{l}$ values only for the rotational phases where the line profiles are completely separated (which means $B_{l}$ values are not computed for overlapping orbital phases, when the system is within 20\% of conjunction). The measured values of $B_{l}$ and their corresponding uncertainties for both components of FK~Aqr are given in columns 5 to 8 of Table~\ref{table:journal}.

\begin{figure}[t!]
    \centering
    \includegraphics[scale=0.5]{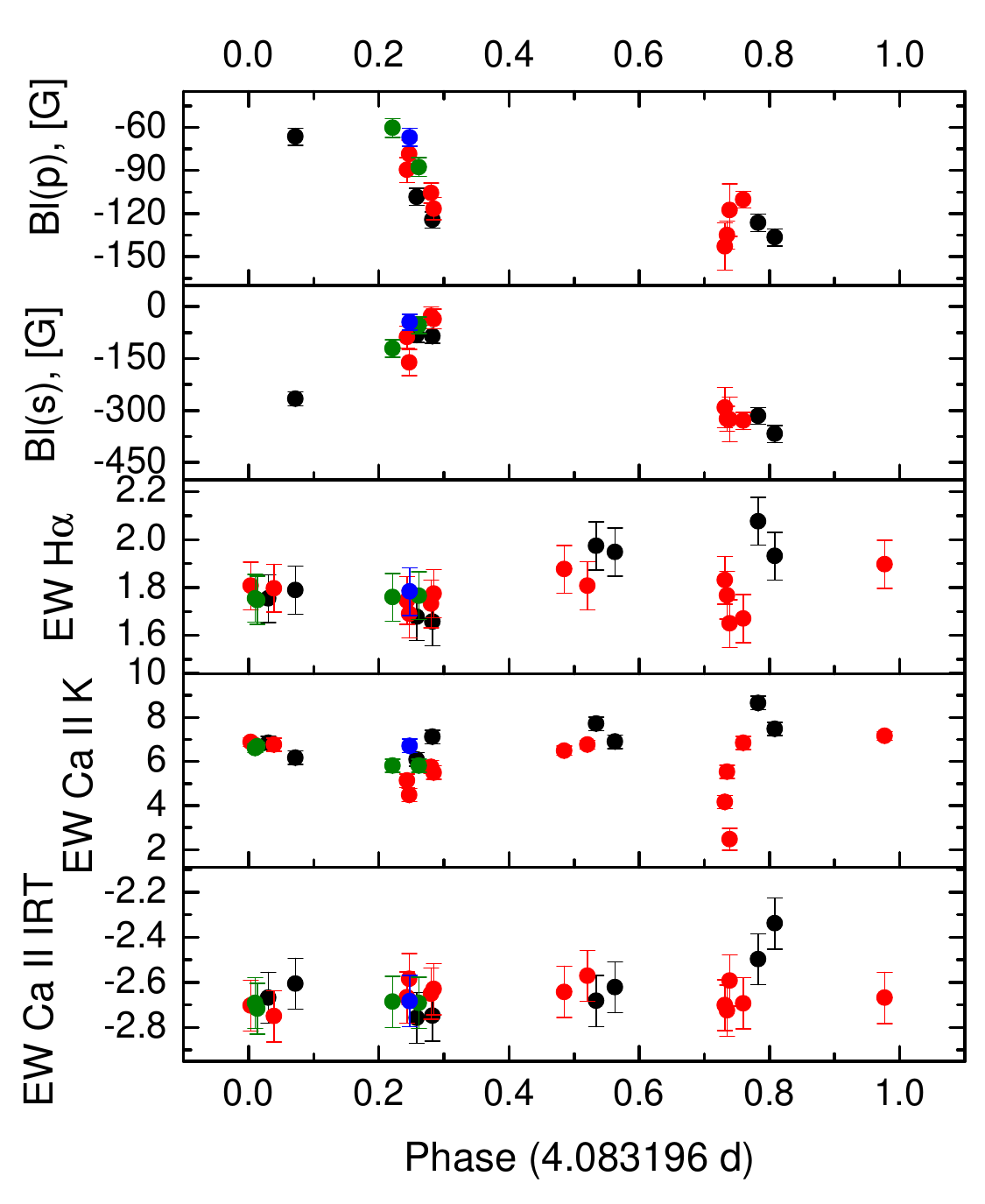}
     \caption{Time variability of the magnetic field and line activity indicators of FK~Aqr. First two plots from top to bottom show the variability of the longitudinal magnetic fields $B_{l}$ of the two components of the system in the period 3--16 September 2014 (where (p) and (s) stand for the primary and secondary, respectively). Observations are phased and the rotational cycles are denoted in black, red, green, and dark blue colours corresponding to the first, second, third, and fourth cycles. The last three plots show the variability of the equivalent widths of the spectral lines, given in {\AA}. All error bars are plotted, but some of them are within the symbols.}
   \label{fig:FKAqrBl}
   \end{figure}

The measured values of $B_{l}$ vary in the interval from $-143 \pm 17$~G to $-60 \pm 7$~G for the primary (denoted (p)) and from $-368 \pm 25$~G to $-27 \pm 26$~G for the secondary (denoted (s)) in Table~\ref{table:journal} and Fig.~\ref{fig:FKAqrBl} (first two plots from top to bottom). The observations are phased according to Eq.~\ref{eq:ephemeris}, which shows four rotational cycles. We coloured them on the plot as follows: the first cycle is denoted in black, the second in red, the third in green (only two observations from 12 September 2014, phases 0.22 and 0.26), and the fourth in dark blue (only one observation from 16 September 2014, phase 0.25).

$B_{l}$ measurements for both stars match well with the reconstructed Zeeman Doppler maps presented in Sect.~\ref{subsec:ZDImaps} and one can easily connect the $B_l$ variation with the main features of these maps.

\subsection{Line activity indicators}

In general, M~dwarfs that are magnetically active and show emission in CaII H\&K and Balmer lines are called `dMe' dwarfs (Dyer 1954). Both components of the binary FK~Aqr belong to this group. We investigated the time variability of the three classical line activity indicators in both components, which are the H${\alpha}$ Balmer line at 656.3~nm, the CaII H\&K lines at 396.8~nm and 393.4~nm, respectively, and the CaII infrared triplet (IRT) at 849.8~nm, 854.2~nm, and 866.2~nm. Before calculating the equivalent widths (EWs) of the lines, we refined the continuum normalisation of the individual orders containing these spectral lines for each spectrum. As there are phases in which the lines of the primary and secondary component are partially or completely blended, we present the sum of the EWs of the two components at all phases. The EWs are computed by integrating over the spectral lines. The resulting EW values in {\AA} are listed in Table~\ref{table:spectral_lines} and displayed on Fig.~\ref{fig:FKAqrBl}.

\subsubsection{H${\alpha}$}

The H${\alpha}$ lines of both components are in emission, as is generally true for the most active stars (Cram \& Mullan 1979). Both components show the typical double-peak (self-reversal) profile for dMe-dwarfs (Worden et al. 1981, Stauffer \& Hartmann 1986), showing that the line is formed in an optically thick chromosphere. This behaviour is well reproduced by chromospheric models with a high temperature gradient and therefore a very high non-thermal heating rate (Houdebine \& Doyle 1994~a, b). Robinson \& Cram (1989) and Robinson et al. (1990) also observed the same type of profile for both components of FK~Aqr. This self-reversal is asymmetric in all our observations, and in general is related to mass motions in the stellar atmosphere. The peak asymmetries are observed in other active single M~dwarfs in both quiescent states and flaring activity (Fuhrmeister et al. 2008, 2011).

\subsubsection{CaII H\&K}

The CaII H\&K lines of both components are in emission. The CaII K lines of the two components are not blended, which is not the case for the CaII H lines. The CaII H line of the secondary is evidently blended with the H$\epsilon$ emission line of the primary (and vice versa, the CaII H line of the primary is blended with the H$\epsilon$ emission line of the secondary, depending on the orbital phase). We therefore present the behaviour of only the CaII K lines.

We noticed significant variation in the EW of the CaII K line from one rotation cycle to the next (Fig.~\ref{fig:FKAqrBl} depicted as black and red dots) near phase 0.73. The four observations corresponding to the second rotation cycle (red dots on Fig.~\ref{fig:FKAqrBl}) were taken consecutively on September 10, 2014,  within a three-hour window. 

The first three observations, which result in discrepant EW values, have the lowest S/Ns for the CaII K order within our data set: 11, 17, and 8 for the first, second, and third observation, respectively (while the fourth observation reaches a S/N of 28). They also correspond to the least precise $B_{l}$ measurements in our data set (see Fig.~\ref{fig:FKAqrBl} top panel). We therefore attribute this strong apparent variability in the CaII K EW on September 10, 2014, to a poorly defined continuum at the blue edge of lower S/N spectra.

\subsubsection{CaII IRT}

The three lines of the CaII IRT (849.8~nm, 854.2~nm, 866.2~nm) of the primary are observed in absorption with emission cores. The emission cores of the lines at 849.8~nm and 854.2~nm reach the continuum and exceed it at all phases in our observations. The emission cores of the third line of the calcium triplet at 866.2~nm are below the continuum (or only just reach it) only in the phases of conjunction. The emission cores of the secondary are superimposed on the wings of the spectral line of the primary for all three lines of the CaII IRT.

\begin{table}

\centering
\begin{center}

\caption{Sum of the equivalent widths of both components of the binary FK~Aqr for the lines H${\alpha}$, CaII H\&K, and CaII IRT (854.2~nm and 866.2~nm). All the equivalent widths are given in {\AA}.}

\label{table:spectral_lines}

\centering
\begin{tabular}{c c c c}
\hline\hline

Date  & EW H${\alpha}$    & EW CaII K         & EW CaII IRT \\
UT    & [$\textup{\AA}$] & [$\textup{\AA}$] & [$\textup{\AA}$]\\
\hline
(1)   & (2)               &  (3)              & (4)\\
\hline
\\
03 Sep 2014 & 1.75 $\pm$ 0.09 & 6.84 $\pm$ 0.26 & -2.67 $\pm$ 0.11 \\
04 Sep 2014 & 1.79 $\pm$ 0.12 & 6.17 $\pm$ 0.27 & -2.61 $\pm$ 0.11 \\
04 Sep 2014 & 1.68 $\pm$ 0.13 & 6.10 $\pm$ 0.30 & -2.76 $\pm$ 0.11 \\
04 Sep 2014 & 1.66 $\pm$ 0.13 & 7.11 $\pm$ 0.27 & -2.75 $\pm$ 0.11 \\
05 Sep 2014 & 1.97 $\pm$ 0.10 & 7.71 $\pm$ 0.29 & -2.68 $\pm$ 0.11 \\
06 Sep 2014 & 1.95 $\pm$ 0.12 & 6.89 $\pm$ 0.27 & -2.62 $\pm$ 0.11 \\
06 Sep 2014 & 2.08 $\pm$ 0.14 & 8.66 $\pm$ 0.30 & -2.50 $\pm$ 0.11 \\
07 Sep 2014 & 1.93 $\pm$ 0.14 & 7.47 $\pm$ 0.29 & -2.34 $\pm$ 0.11 \\
07 Sep 2014 & 1.81 $\pm$ 0.09 & 6.88 $\pm$ 0.21 & -2.70 $\pm$ 0.11 \\
07 Sep 2014 & 1.80 $\pm$ 0.10 & 6.75 $\pm$ 0.25 & -2.75 $\pm$ 0.11 \\
08 Sep 2014 & 1.75 $\pm$ 0.13 & 5.12 $\pm$ 0.28 & -2.67 $\pm$ 0.11 \\
08 Sep 2014 & 1.69 $\pm$ 0.13 & 4.48 $\pm$ 0.27 & -2.59 $\pm$ 0.11 \\
08 Sep 2014 & 1.73 $\pm$ 0.13 & 5.75 $\pm$ 0.28 & -2.65 $\pm$ 0.11 \\
08 Sep 2014 & 1.77 $\pm$ 0.13 & 5.50 $\pm$ 0.30 & -2.63 $\pm$ 0.11 \\
09 Sep 2014 & 1.88 $\pm$ 0.08 & 6.49 $\pm$ 0.24 & -2.64 $\pm$ 0.11 \\
09 Sep 2014 & 1.81 $\pm$ 0.08 & 6.77 $\pm$ 0.22 & -2.57 $\pm$ 0.11 \\
10 Sep 2014 & 1.83 $\pm$ 0.14 & 4.16 $\pm$ 0.29 & -2.70 $\pm$ 0.11 \\
10 Sep 2014 & 1.77 $\pm$ 0.14 & 5.54 $\pm$ 0.30 & -2.73 $\pm$ 0.11 \\
10 Sep 2014 & 1.65 $\pm$ 0.14 & 2.47 $\pm$ 0.46 & -2.59 $\pm$ 0.11 \\
10 Sep 2014 & 1.67 $\pm$ 0.14 & 6.85 $\pm$ 0.28 & -2.69 $\pm$ 0.11 \\
11 Sep 2014 & 1.90 $\pm$ 0.09 & 7.15 $\pm$ 0.22 & -2.67 $\pm$ 0.11 \\
11 Sep 2014 & 1.76 $\pm$ 0.09 & 6.61 $\pm$ 0.21 & -2.69 $\pm$ 0.11 \\
11 Sep 2014 & 1.75 $\pm$ 0.09 & 6.68 $\pm$ 0.20 & -2.72 $\pm$ 0.11 \\
12 Sep 2014 & 1.76 $\pm$ 0.13 & 5.81 $\pm$ 0.28 & -2.69 $\pm$ 0.11 \\
12 Sep 2014 & 1.77 $\pm$ 0.13 & 5.80 $\pm$ 0.29 & -2.69 $\pm$ 0.11 \\
16 Sep 2014 & 1.78 $\pm$ 0.13 & 6.71 $\pm$ 0.28 & -2.68 $\pm$ 0.11 \\

\\

\hline
\hline
\end{tabular}
\end{center}
\end{table}

\section{Zeeman Doppler imaging}
\label{subsec:ZDImaps}

We used the ZDI method (Semel 1989, Donati \& Brown 1997, Donati et al. 2006~b) to reconstruct the surface magnetic field topologies of both components of the binary FK~Aqr. This method uses the rotation-induced modulation and Doppler shifts of Stokes $V$ signatures to map the surface vector magnetic field decomposed onto a poloidal--toroidal spherical harmonics frame (Donati et al. 2006~b). The iterative algorithm fits the observed LSD Stokes profiles with a set of simulated profiles corresponding to the same rotational phases. These synthetic Stokes profiles are computed from a model star, the surface of which is divided into a grid of 2000 pixels of roughly the same area. ZDI uses maximum entropy regularisation as described by Brown et al. (1991), who implement the algorithm for maximum entropy optimisation developed by Skilling \& Bryan (1984). The local synthetic Stokes $I$ line profile is assumed to possess a Voigt shape. The method also uses the weak-field assumption (see e.g. Donati et al. 2003).

In the present paper, we use a version of the code \textsc{ZDIpy}\footnote{https://github.com/folsomcp/ZDIpy} developed by Folsom et al. (2018) that has been adapted to  simultaneously map both components of an SB2 spectroscopic binary system (referred to as the SB version of the code hereafter). \textsc{ZDIpy} is Python-based and implements the same physical model, analysis principles, and coefficient definition as the code of Donati et al. (2006~b) described above. We considered spherical stars in the present study. This is well-suited given the high value of the ratio $\frac{a_A+a_B}{R_A+R_B}=17$ (the binary is non-interacting in the case where the ratio is greater than 10; e.g. Eker et al. 2014). The distance of the Lagrangian point $L_1$ from the primary of FK~Aqr was calculated to be around $0.53~a$ when employing the classical formula given by Lagrange (where $a$ is the separation between the two stars; Coel Hellier 2001, Leahy \& Leahy 2015), showing no departure from sphericity. A second source of non-sphericity in general could be the stellar rotation itself. The criterion for this is given by Cang et al. (2020, their equation (3)); the magnitude of the oblateness is given by the relation between the ratio $R_p/R_e$ (polar and equatorial radii, respectively) and the stellar rotation rate $\Omega$. We checked this criterion for FK~Aqr and conclude that it too is negligible ($R_p/R_e \approx 1$).

The method requires several input parameters. The adopted values for the projected rotational velocities of FK~Aqr are $v\sin i = 5.4 \pm 0.6$~km\,s$^{-1}$ for the primary and $v\sin i = 4.4 \pm 0.6$~km\,s$^{-1}$ for the secondary (both values calculated in section~\ref{subsec:vsini_values}). We assume synchronous rotation; the adopted value for both rotation periods is therefore $P_{\rm rot}=P_{\rm orb}=4.08319598 \pm 0.00000095$~d, which is the output value of our \textsc{Phoebe} analysis. According to the Zahn (1977) formalism, we indeed compute a synchronisation timescale of shorter than 50~Myr for FK~Aqr. The inclination angle of the binary is set to $52.39^{\circ} \pm 0.45^{\circ}$ (derived from the astrometric analysis described in Sect.~\ref{subsec:astrometry}) and we assume the stellar rotation axes are perpendicular to the orbital plane. The spherical harmonics expansion is limited to $\ell = 10$, a value well suited to the moderate $v\sin i$ values. The linear limb darkening coefficients for both components of FK~Aqr are set to 0.72 according to Claret (2004). 

The local line profile parameters (line strength, Gaussian, and Lorentzian widths) are adjusted to achieve the best fit between synthetic and observed Stokes~$I$ line profiles. Although the line parameters are partly degenerate, our tests show that the reconstructed magnetic maps are weakly sensitive to the precise choice of a set of line parameters (see Appendix~\ref{appendix:ZDI_models}).

Another input parameter required by the SB version of \textsc{ZDIpy} is the flux ratio of the two components of the binary. This parameter is not available in the literature. We therefore employed the maximum entropy method to fit the models with observations by keeping all the parameters
except the flux ratio at same values. In this way, the model with the maximum value of entropy (i.e. the weakest magnetic field) gave us a value for the flux ratio equal to F$_s$/F$_p$ = 0.3\footnote{This value corresponds to an S/N-weighted average value on the ESPaDOnS spectral domain, roughly corresponding to the $R$- or $I$-band. As expected for a secondary with later spectral type than the primary, we find a value lower than that derived in Sect.~\ref{subsec:pionier} and corresponding to the $H$-band.}, which we use in our final model. Finally, the SB version of \textsc{ZDIpy} also requires the radial velocities of both components at each observed rotational phase. We used the RV values derived from our \textsc{Phoebe} orbital model.

\begin{figure*}
    \centering
    \includegraphics[]{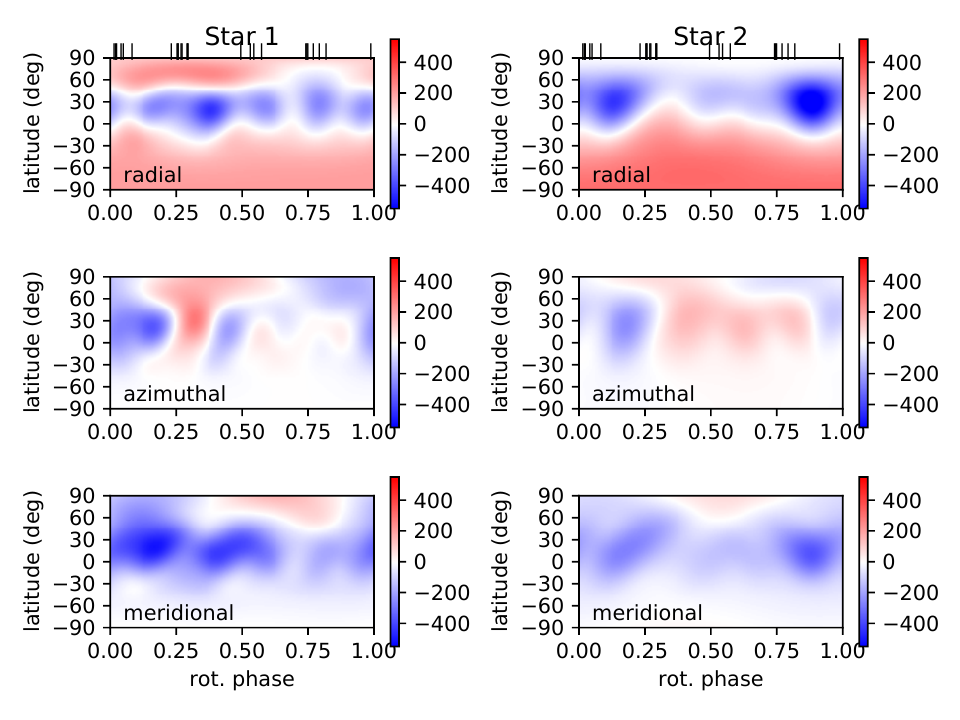}
     \caption{Magnetic maps of both components of the system FK~Aqr (primary on the left and secondary on the right) for the observational period 3--16 September 2014. From top to bottom, we show the field components in spherical coordinates: radial, azimuthal, and meridional. On the right side of each subplot, we provide a colour bar showing the magnetic field strength expressed in gauss. The phases of the observations are marked on top of each radial map.}
   \label{fig:FKAqrZDI}
   \end{figure*}

Our observations span four rotational cycles and provide a reasonably even and dense phase coverage. The ZDI model is almost able to fit our data set down to the noise level with a reduced $\chi_{r}^2$ of 1.7. As can be seen in Fig.~\ref{fig:FKAqrStokesIV}, the model accurately reproduces the modulation of the high-quality Stokes~$V$ data along the stellar rotation and orbit. The reconstructed large-scale magnetic field topologies of the two components are presented in Fig.~\ref{fig:FKAqrZDI}. The properties of the surface magnetic field of both stars are very similar: an average magnetic field modulus $B_{mean}\simeq 250$~G, and the magnetic topologies are almost purely poloidal and dipole-dominated. The statistics of the magnetic fields of both components are presented in Table~\ref{table:magnetic_properties}.

The radial magnetic map of the primary features a positive polar region in the hemisphere orientated towards the observer, which is otherwise dominated by a negative polarity (Fig.~\ref{fig:FKAqrZDI}). This feature translates into a rather strong quadrupolar component ($\simeq 22~\%$ of the reconstructed magnetic energy) for this star. This polar region of opposite polarity may seem surprising given the simple two-lobe Stokes V profiles observed at all phases in Fig.~\ref{fig:FKAqrStokesIV}. We therefore further investigated the sensitivity of the appearance of this feature to various ZDI inputs. First, we explored the map evolution as a function of the target $\chi_{r}^2$ for our ZDI model. We found that this polar region of positive polarity weakens with increasing target $\chi_{r}^2$ and completely disappears for $\chi_{r}^2$ above 2.5 (see Appendix~\ref{appendix:ZDI_model_chi2}). Second, we investigated the effect of the spectra taken at phases during which the spectral lines of the two stars overlap. As shown in Appendix~\ref{appendix:ZDI_model_no_overlap}, when removing these spectra from our model, the polar region of positive polarity weakens and correspondingly the fraction of magnetic energy residing in the quadrupolar component decreases. This shows that this reverse polarity pole is partly due to cross-talk between the magnetic maps of the two components. Finally, we note that our $v\sin i$ values are based on radii derived from stellar models suited to inactive stars, whereas several studies of active M~dwarfs infer a significant radius inflation (e.g. \cite{Lopez-Morales07},  Donati et al. 2008). We therefore tested the effect of varying our input $v\sin i$ values, and found that the polar region of positive polarity disappears when increasing the $v\sin i$ value above 6.4~km\,s$^{-1}$ for the primary (see Appendix~\ref{appendix:ZDI_model_vsini_inflated}). Although this value would correspond to an unrealistically large radius inflation of 30~\% with respect to single M~dwarfs, our tests confirm that the presence of this high-latitude area of positive polarity on the primary component of FK~Aqr cannot be considered a reliable feature.

\begin{table*}

\centering
\begin{center}

\caption{Magnetic analysis of the components of FK~Aqr.}

\label{table:magnetic_properties}

\centering
\begin{tabular}{c c c c c c c c c}
\hline\hline
component & $B_{mean}$ & $B_{max}$ & poloidal & toroidal & dipole   & quadrupole & octupole & axisymmetric \\
          &  [G]       &   [G]      & [\% tot] & [\% tot] & [\% pol] & [\% pol]  & [\% pol] & [\% tot]     \\
\hline
\\
primary   & 248.8 & 675.4 & 89.6 & 10.4 & 56.8 & 22.1 & 13.0 & 72.1 \\
secondary & 250.8 & 702.8 & 95.9 & 4.1 & 78.5 & 9.6 & 7.5 & 70.4 \\
\\

\hline
\hline
\end{tabular}
\end{center}
\end{table*}

 Glebocki \& Stawikowski (1995) conclude that the possibility of a different inclination of the primary of FK~Aqr to the orbital plane  is highly unlikely. Nevertheless, we ran some new models, changing the inclination angle of the primary. Small changes in inclination have almost no effect. Even for very large changes in inclination, the major features of the map are largely unchanged. Additionally, we ran the single-star version of ZDIpy with the Unno-Rachkovsky implementation (Bellotti et al. 2023) on the non-overlapping profiles of the primary. This model shows almost no difference compared to the model without Unno-Rachkovsky implementation. Unno-Rachkovsky model implementation in the ZDIpy binary-star version is deferred to a future paper.

\section{Summary and discussion}
We conducted an observing campaign targeting the M~dwarf binary system FK~Aqr in the framework of the BinaMIcS project (Binary and Magnetic Interactions in various classes of Stars, Alecian et al. 2015~a, b) with the aim being to study stellar magnetism under the influence of the physical processes occurring in close binaries. Combining our astrometric and radial velocity measurements with archival radial velocities, we refined the orbital parameters of the system and derived an orbital inclination of the system of $52.39^{\circ} \pm 0.45^{\circ}$ and deprojected masses of the components of $M_1=0.55 \pm 0.01~{\rm M_{\odot}}$ and $M_2=0.44 \pm 0.01~{\rm M_{\odot}}$.

This puts both stars just above the so-called full-convection threshold, which means that their internal structure comprises an inner radiative core below a deep convective envelope. The two components of the FK~Aqr system are therefore particularly interesting targets for studying how the transition from partly convective to fully convective stellar dynamos occurs in close binary systems.

The system FK~Aqr was observed with the spectropolarimeter ESPaDOnS (Donati et al. 2006) and 26 spectra in total were collected in the period from  3 to 16 September 2014. We used the least-squares deconvolution (LSD) multi-line technique (Donati et al. 1997) to generate the mean Stokes~$I$ and $V$ line profiles, from which clear Stokes~$V$ signatures are visible from all observations. The signatures have simple shapes with negative blue and positive red lobes. We computed the line-of-sight component of the stellar magnetic field $B_{l}$  using the first-order moment method (Rees \& Semel 1979, Donati et al. 1997,  Wade et al. 2000~a,b), and find it to vary in the interval from -143~G to -60~G for the primary and from -368~G to -27~G for the secondary.

The time variability of the three classical line activity indicators (H${\alpha}$, CaII H\&K, and CaII IRT) of both components was compared to the variability of $B_{l}$. The temporal evolution of the  EW of the three lines is of limited amplitude and varies almost within the error bars. In particular, we do not find evidence for rotational modulation or enhanced activity at specific orbital phases.

We used the Zeeman Doppler imaging tomographic method (ZDI; Semel 1989, Donati \& Brown 1997, Donati et al. 2006~b) to reconstruct the large-scale component of the surface magnetic fields of both M~dwarfs. We used the \textsc{ZDIpy} code presented in Folsom et al. (2018), which is Python-based and adapted to binary stars. The two components host large-scale magnetic fields with similar properties. Both are largely dominated by the poloidal component (with 90~\% and 96~\% of the reconstructed magnetic energy for the primary and secondary, respectively) and are mainly axisymmetric (72~\% and 70~\% for the primary and secondary), featuring a major contribution of the $\ell=1$ poloidal dipole modes (57~\% for the primary and 78~\% for the secondary). The reconstructed magnetic map of the primary also features a significant quadrupolar component (22~\% of the reconstructed magnetic energy), but our tests show that this feature is sensitive to model parameters and to cross-talk between the primary and secondary at conjunction phases, and it could be significantly weaker (down to 12~\%). The mean large-scale magnetic field of both components is close to 250~G, and the local field modulus reaches values close to 700~G in both cases.

Previous studies of the large-scale magnetic field of single rapidly rotating M~dwarfs have revealed the existence of a change in the observed magnetic properties at a mass of approximately 0.4--0.5~M$_\odot$, which is just above the full-convection threshold (Donati et al. 2008, Morin et al. 2008). Stars with saturated activity (corresponding to rotation periods of shorter than $\simeq$5~d) that are more massive than this limit exhibit large-scale magnetic fields of moderate intensity ($B_{\rm mean}$ in the range 100--200~G), a significant or even dominant toroidal component (more than 20~\% of the reconstructed magnetic energy), and often a significant contribution of non-axisymmetric poloidal modes. M~dwarfs just below this mass limit display stronger large-scale magnetic fields (400--800~G) that are almost purely poloidal, are dominated by the dipole mode (more than 60~\% of the reconstructed magnetic energy), and in most cases are strongly axisymmetric. The active M3 dwarf AD~Leo ($M_\star=0.41~{\rm M_\odot}$, $P_{\rm rot}=2.23$~d) can be considered as an intermediate case: it hosts an almost purely poloidal large-scale magnetic field dominated by the axial dipole, but the average field modulus takes intermediate values in the range 200--300~G (Morin et al. 2008, Lavail et al. 2018, Bellotti et al. 2023). The secondary component of FK~Aqr, which has stellar parameters very similar to those of AD~Leo ($M_\star=0.44~{\rm M_\odot}$, $P_{\rm rot}=4.08$~d), generates a large-scale magnetic field of the same type (dipole-dominated of intermediate strength). The case of the primary appears more interesting: its large-scale magnetic field can also be classified as AD~Leo type, although it is significantly more massive ($M_\star=0.55~{\rm M_\odot}$), and is indeed the most massive M~dwarf known to host a dipole-dominated magnetic field of intermediate strength. There are stars with similar parameters in the single M~dwarf sample from Donati et al. (2008), namely OT~Ser ($M_\star=0.55~{\rm M_\odot}$, $P_{\rm rot}=3.40$~d) and DT~Vir ($M_\star=0.59~{\rm M_\odot}$, $P_{\rm rot}=2.85$~d), both of which host a clearly different type of magnetic field, which is significantly weaker and features in particular a strong toroidal component. Interestingly, the young M~dwarf AU~Mic (22~Myr), with mass and rotation period close to those of the FK~Aqr primary ($M_\star=0.50~{\rm M_\odot}$, $P_{\rm rot}=4.84$~d), hosts a significantly stronger ($B_{mean}=475$~G) magnetic field featuring significant toroidal and non-axisymmetric components, although its young age renders a direct comparison less relevant (\cite{Klein2021, Donati23}).

The primary component of the FK~Aqr system represents the high-mass end of the AD~Leo-type of magnetism. However, the role of its binary nature is not yet clear. A valuable point of comparison is provided by the short-period binary YY~Gem, which is composed of two M~dwarfs of almost equal mass ($M_\star=0.61~{\rm M_\odot}$, $P_{\rm rot}=0.81$~d).  For
both components, Kochukhov \& Shulyak (2019) recover a large-scale magnetic field of intermediate strength ($B_{mean}=205$ and 260~G) featuring a significant toroidal component ($\sim 30~\%$ of the magnetic energy), and with roughly half of the magnetic energy in non-axisymmetric modes; that is, the same type of magnetism as single M~dwarfs with similar parameters. Simultaneously, in both systems FK~Aqr and YY~Gem, the primary and the secondary (with similar and almost equal masses, respectively) generate surface magnetic fields with very similar properties. Interestingly, in FK~Aqr, the dipolar components of the magnetic fields of the two stars appear to be aligned, while they are anti-aligned in the case of YY~Gem. 

The system FK~Aqr also provides new constraints on the extent of the parameter domain in which rapidly rotating M~dwarfs are able to generate two different types of large-scale magnetic field: strong dipolar or weaker multipolar fields. \cite{Morin2010} report this behaviour for six stars less massive than $\sim 0.2~{\rm M_\odot}$ with rotation periods of shorter than $\sim2$~d. This could be explained by an effect of age, long-term magnetic cycles (\cite{Kitchatinov2014}), or dynamo bistability (\cite{Morin2011}, \cite{Gastine2013}). This behaviour is also reported by \cite{Kochukhov17} for the two components of the coeval wide binary system GJ~65~AB, which lie in the previously identified domain with masses close to $0.12~{\rm M_\odot}$ (\cite{Kervella2016}) and rotation periods of $\sim 0.25$~d. The secondary UV~Ceti  hosts a strong dipole-dominated field, while the large-scale magnetic field of the primary BL~Ceti is much weaker and less axisymmetric. The pair FK~Aqr provides further evidence that the parameter space where the two types of magnetism co-exist does not extend above 0.2~M$_\odot$, at least for stars in close binary systems.

Future analyses of other M~dwarf binary systems observed in the framework of the BinaMIcS project, combined with new results on the magnetism of single M~dwarfs ---in particular based on observations collected with the near-infrared spectropolarimeter SPIRou--- will contribute to disentangling the effects of mass, rotation, age, and binarity on the dynamo-generated magnetism of main sequence stars close to the full-convection threshold.

\begin{acknowledgements}
      S.Ts. acknowledges the funding from CNRS/IN2P3. S.Ts. is thankful to Kyle Conroy, one of the developers of the \textsc{Phoebe} code, for the priceless discussions about the code itself. The authors thank the CFHT staff for their valuable help throughout the BinaMIcS large program. GAW acknowledges Discovery Grant support from the Natural Sciences and Engineering Research Council (NSERC) of Canada. JM, JBLB, EA, SB, CN, and PP acknowledge support from the "Programme National de Physique Stellaire" (PNPS) of CNRS/INSU co-funded by CEA and CNES. This work has been partially supported by a grant from Labex OSUG@2020 (Investissements d’avenir – ANR10 LABX56)
\end{acknowledgements}

%
%

\onecolumn

\begin{appendix}

\section{A combined fit of the radial velocities and the astrometric positions}  
\label{appendix:astrometry_and_RV}

We ran a combined fit of the radial velocities and the astrometric positions using the code presented in \cite{LeBouquin2017}. We find that the radial velocities of HM65 are systematically offset with respect to the best-fit solution, which is mostly driven by the new observations. We assume that this difference can be explained by the accuracy of the HM65 radial velocity measurements, which are based on chromospheric emission lines and shift their values by +1.4~km\,s$^{-1}$. The best-fit parameters are summarised in Table~\ref{table:best_fit_astrometry_and_rv}.

The eccentricity is very small but significantly different from zero. We find that this small residual eccentricity is entirely driven by the new ESPaDOnS radial velocities, but is not detected when using only the historical radial velocities and the astrometric observations.

Combining the radial velocity amplitudes, the period, and the apparent size of the astrometric orbit, it is possible to estimate the geometrical distance to the system and the individual masses of the two components. The inferred distance of $d=8.876\pm0.098$~pc from the combined fit is compatible with the independent estimation of $d=8.897\pm0.004$~pc from Gaia parallax.

\begin{table*} [h!]
\centering
\caption{ Best-fit parameters of a combined fit of the radial velocities and the astrometric observations.}
\label{table:best_fit_astrometry_and_rv}
 \begin{tabular*}{6.35cm}{cccc}
 \hline\hline\noalign{\smallskip}
 Element & Unit & Value & Uncertainty \\
 \noalign{\smallskip}\hline\noalign{\smallskip}
 $T$ & MJD  &  $37145.075$  &  $0.037$  \\
 $P$ & days  &  $4.0831962$  &  $0.0000029$  \\
 $a$ & mas  &  $5.607$  &  $0.036$  \\
 $e$ &   &  $0.00520$  &  $0.00047$  \\
 $\Omega$ & deg  &  $111.64$  &  $0.22$  \\
 $\omega$ & deg  &  $307.5$  &  $3.0$  \\
 $i$ & deg  &  $51.95$  &  $0.42$  \\
 $K_a$ & km/s  &  $46.138$  &  $0.011$  \\
 $K_b$ & km/s  &  $58.275$  &  $0.023$  \\
 $g$ & km/s  &  $-7.3758$  &  $0.0088$  \\
 \noalign{\smallskip}\hline\noalign{\smallskip}
 \multicolumn{4}{c}{From apparent orbit and radial velocities}\\
 $d$ & pc  &  $8.876$  &  $0.098$  \\
 $M_a$ & ${M}_\odot$  &  $0.5503$  &  $0.0095$  \\
 $M_b$ & ${M}_\odot$  &  $0.4357$  &  $0.0075$  \\
 \noalign{\smallskip}\hline
 \end{tabular*}
\end{table*}


\section{ZDI models}  
\label{appendix:ZDI_models}

In order to create the magnetic models, we need a set of three parameters to describe the mean Stokes $I$ line: the strength, Gauss, and Lorentz widths. We used a script to optimise the line model parameters based on a $\chi_{r}^2$ minimisation of the median profile. We searched in the interval between 0.5 and 5.0 with a step of 0.1 for all the three parameters. We decided to examine the outcomes of two approaches: The first is to keep the three parameters equal for both stars, which could be expected because they belong to one and the same spectral class. The second is to allow the script to search for different parameters of the primary and secondary. This resulted in many combinations that fit the lines well. The $\chi_{r}^2$ per spectrum of all the fits varies between 0.21 and 1.0 with the tendency for $\chi_{r}^2$ to be larger in the case of equal parameters for the primary and secondary than the case when the parameters are different for the two stars. ZDI models were reconstructed with all these sets of line parameters. Their $\chi_{r}^2$ vary between 1.5 and 2.1. The magnetic field structures of the two stars as seen in Fig.~\ref{fig:FKAqrZDI} remain the same for all of the more than 50 models. Changing the sets of the three line parameters led to only small changes in the magnetic field properties.

In order to select the best models, we decided to limit the $\chi_{r}^2$ of the Stokes $I$ line fits following the rule $\chi_{r}^2 + 1/n$, where n is the degrees of freedom. In the table below, Table~\ref{table:models}, we present the variability of the magnetic characteristics from all of the models. We consider that these results can be used as error bars.

\begin{table*} [h!]

\centering
\begin{center}

\caption{Magnetic analysis from all of the ZDI models.}

\label{table:models}

\centering
\begin{tabular}{c c c c c c c c c}
\hline\hline
component & $B_{mean}$ & $B_{max}$ & poloidal & toroidal & dipole   & quadrupole & octupole & axisymmetric \\
          &  [G]       &   [G]      & [\% tot] & [\% tot] & [\% pol] & [\% pol]  & [\% pol] & [\% tot]     \\
\hline
\\
primary & 224--249 & 539--675 & 90--92 & 8--10 & 57--62 & 20--23 & 12--13 & 72--79 \\
secondary & 230--269 & 537--703 & 96--99 & 1--4 & 79--92 & 4--10 & 3--8 & 70--76 \\
\\

\hline
\hline
\end{tabular}
\end{center}
\end{table*}


\section{ZDI model with a higher $\chi_{r}^2$}  
\label{appendix:ZDI_model_chi2}

\begin{figure*} [h!]
    \sidecaption
    \includegraphics[width=0.3\textwidth]{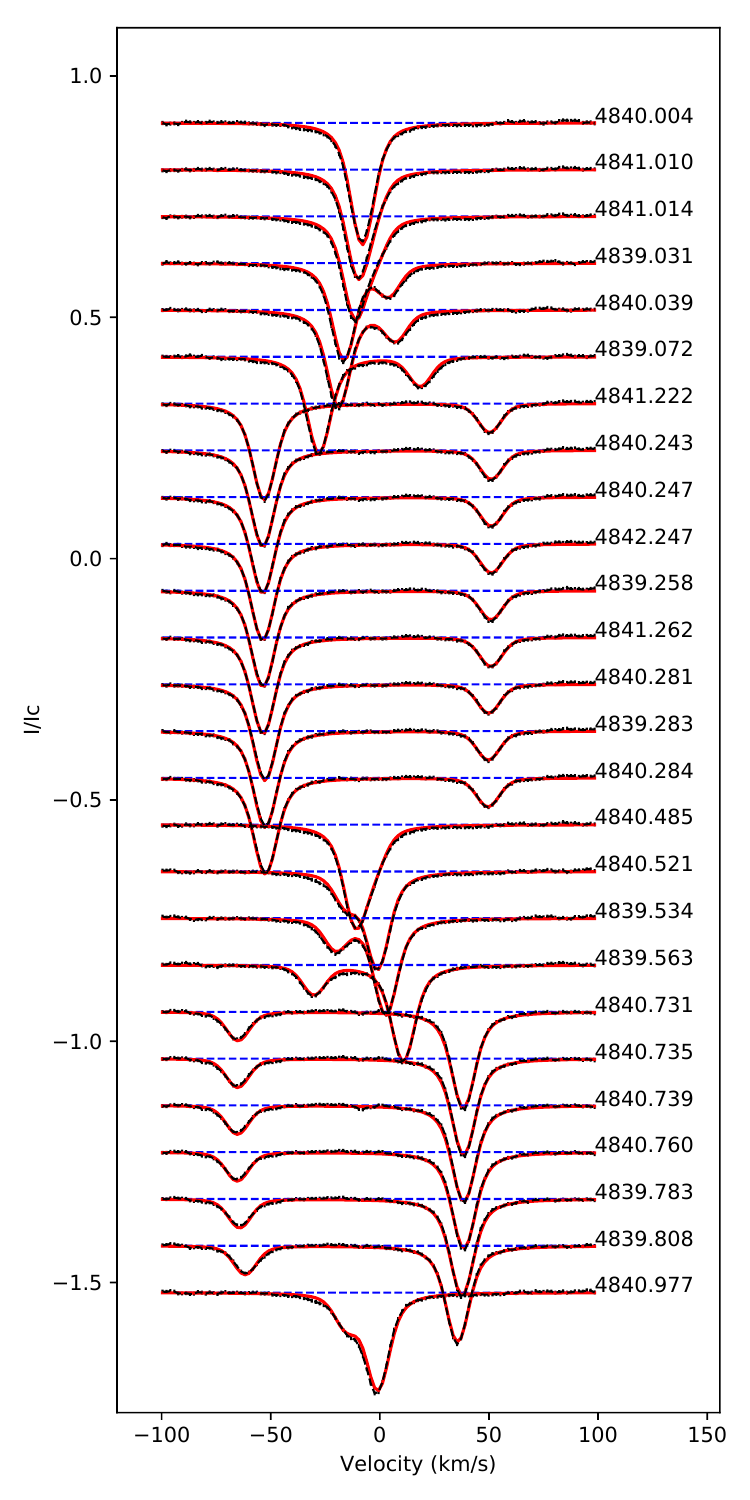}
    \includegraphics[height=0.45\textheight]{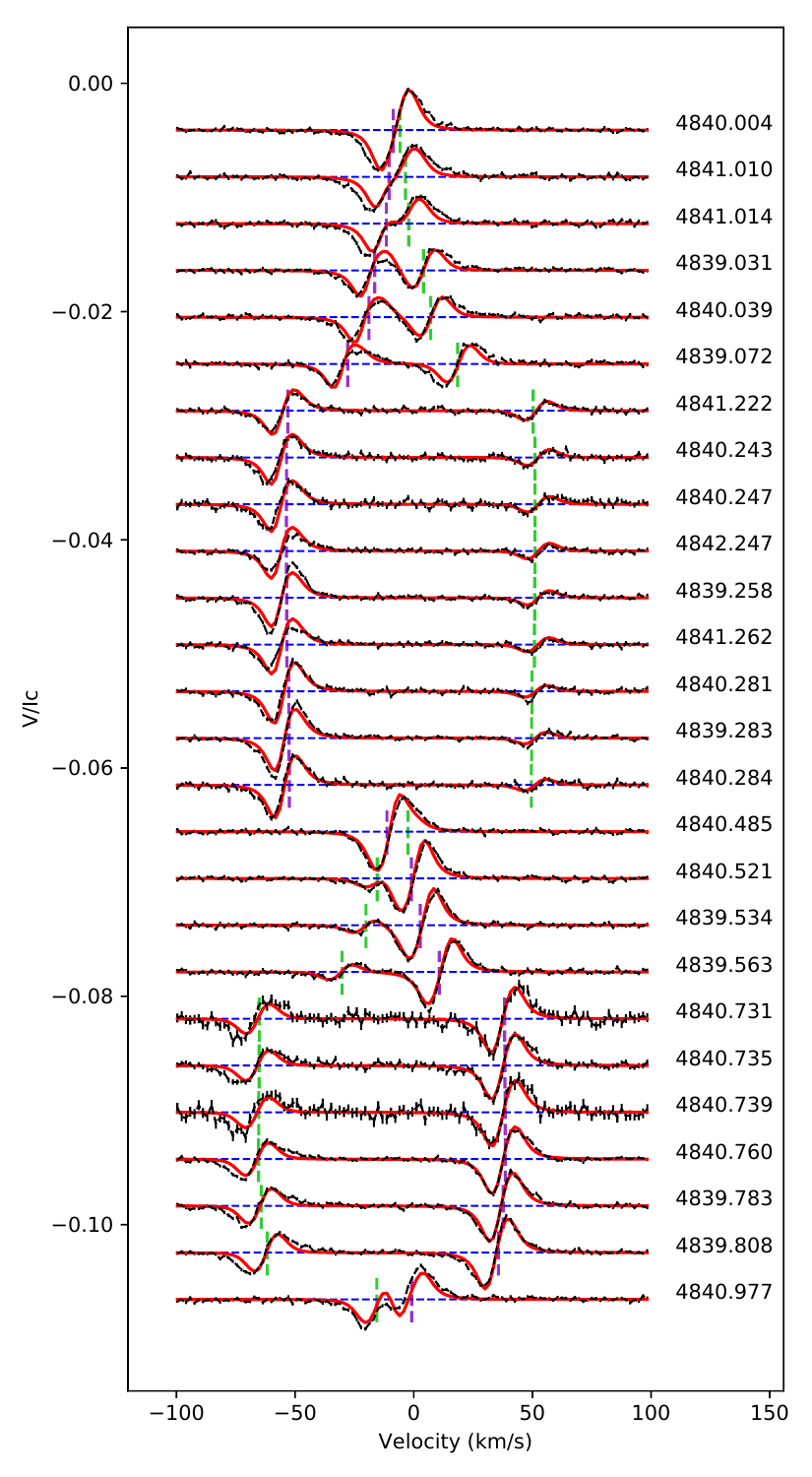}
     \caption{Normalised Stokes $I$ and $V$ profiles of FK~Aqr in the case of a model with $\chi_{r}^2 = 2.5$.}
   \label{fig:stokesIV_appendix_higher_chi2}
   \end{figure*}

\begin{figure*} [h!]
    \centering
    \includegraphics[]{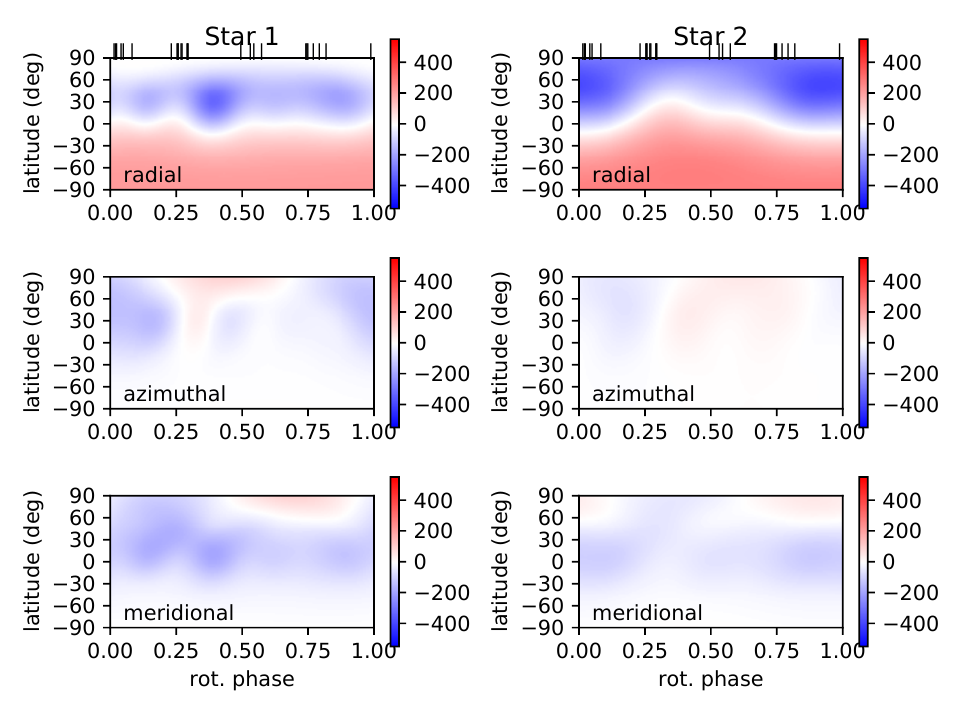}
     \caption{Magnetic maps of both components of the system FK~Aqr in the case of a model with $\chi_{r}^2 = 2.5$.}
   \label{fig:zdi_appendix_higher_chi2}
   \end{figure*}

Here, we present a ZDI model reconstructed with all observations of both stars and all the parameters are kept with the same values as the main model. We only gradually increase (with a step of 0.1) the $\chi_{r}^2$ value starting from 1.7 to examine the behaviour of the polar region of positive polarity for the primary. We noticed that the higher the $\chi_{r}^2$ value, the weaker this region is; above $\chi_{r}^2 = 2.5,$ it disappears (Fig.~\ref{fig:stokesIV_appendix_higher_chi2} and Fig.~\ref{fig:zdi_appendix_higher_chi2}). The magnetic characteristics of the model are given in Table~\ref{table:chi2_25}.


\begin{table*} [h!]

\centering
\begin{center}

\caption{Magnetic analysis in the case of a model with $\chi_{r}^2 = 2.5$.}

\label{table:chi2_25}

\centering
\begin{tabular}{c c c c c c c c c}
\hline\hline
component & $B_{mean}$ & $B_{max}$ & poloidal & toroidal & dipole   & quadrupole & octupole & axisymmetric \\
          &  [G]       &   [G]      & [\% tot] & [\% tot] & [\% pol] & [\% pol]  & [\% pol] & [\% tot]     \\
\hline
\\
primary   & 165.1 & 361.6 & 92.8 & 7.2 & 81.3 & 9.3 & 6.9 & 87.3 \\
secondary & 188.1 & 399.5 & 99.9 & 0.1 & 95.2 & 3.5 & 1.2 & 82.7 \\
\\
\hline
\hline
\end{tabular}
\end{center}
\end{table*}

\vspace{5cm}

\section{A model without phases with overlapping line profiles} 
\label{appendix:ZDI_model_no_overlap}

\begin{figure*} [h!]
    \sidecaption
    \includegraphics[width=0.27\textwidth]{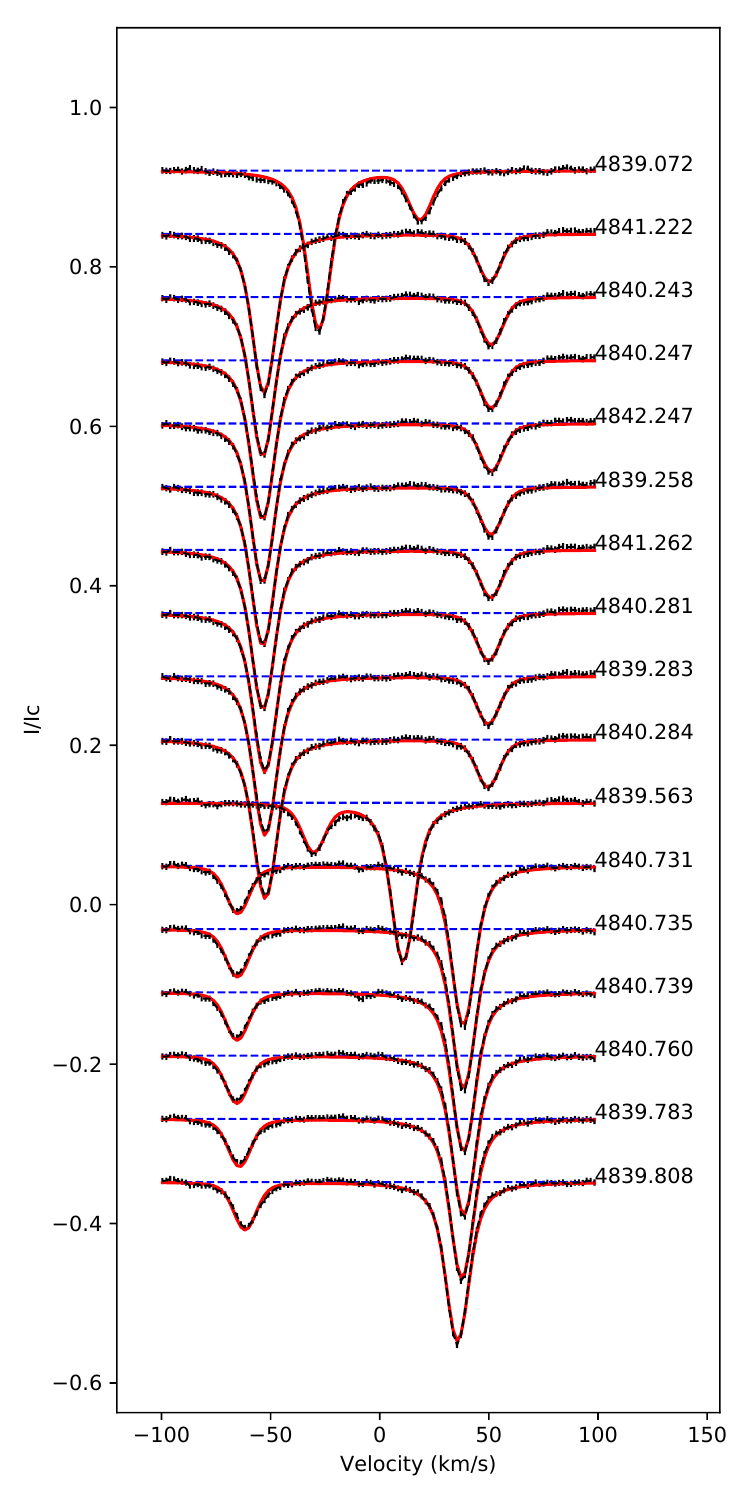}
    \includegraphics[height=0.40\textheight]{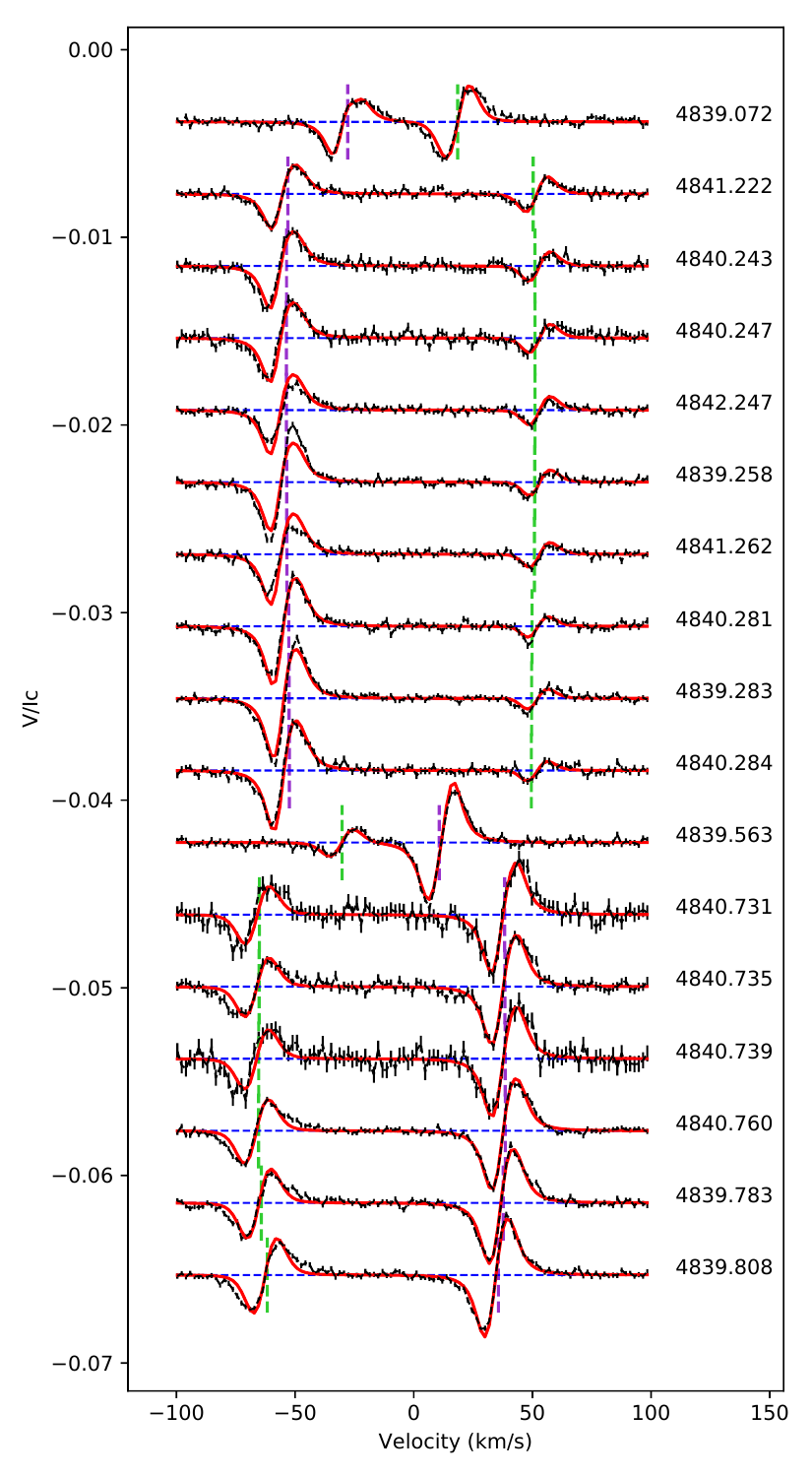}
     \caption{Normalised Stokes $I$ and $V$ profiles of FK~Aqr in the case using only the phases where the line profiles of the two components do not overlap.}
   \label{fig:stokesIV_nooverlap}
   \end{figure*}

\begin{figure*} [h!]
    \centering
    \includegraphics[]{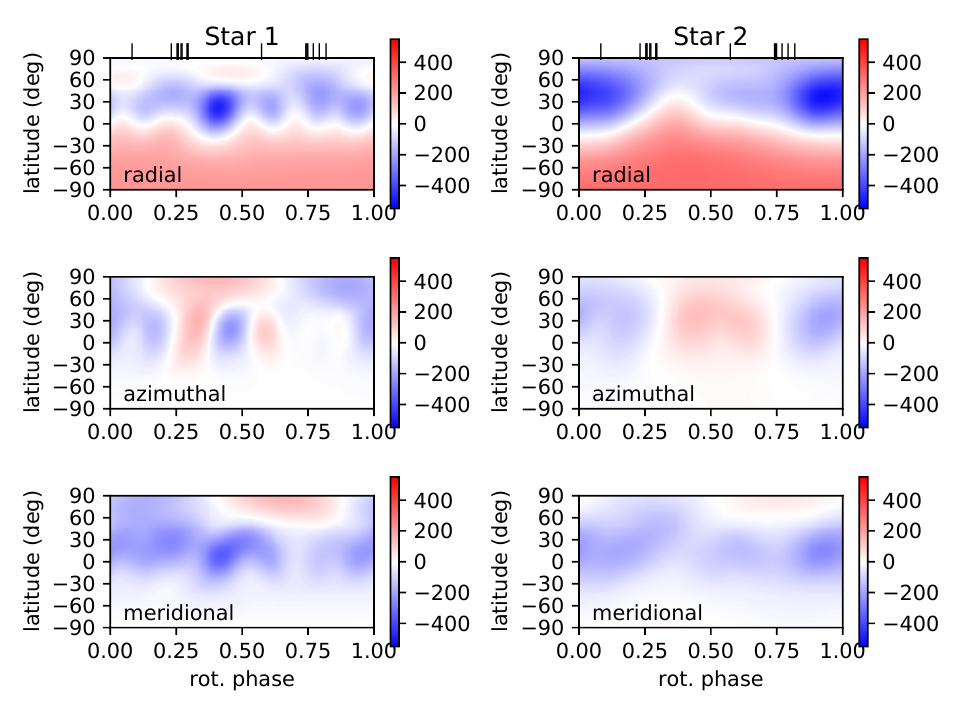}
     \caption{Magnetic maps of both components of the system FK~Aqr in the case using only the phases when the line profiles of the two components do not overlap.}
   \label{fig:zdi_nooverlap}
   \end{figure*}

We explore the case in which the magnetic field topologies of both stars are reconstructed using only the phases where the line profiles do not overlap. Thus, we use only 17 observations out of 26 in total. We kept all the parameters for this model equal to the parameters of the main model presented in the main text. As a result, the $\chi_{r}^2$ is 1.6 compared to 1.7 of the main model. The magnetic characteristics of the model are given in Table~\ref{table:less_profiles}. The corresponding fit of Stokes $I$ and $V$ profiles and the magnetic maps are given in Fig.~\ref{fig:stokesIV_nooverlap} and Fig.~\ref{fig:zdi_nooverlap}. The large-scale magnetic field structure is not affected by the reduction of observations and remains the same. The magnetic field strength is slightly weaker.


\begin{table*} [h!]

\centering
\begin{center}

\caption{Magnetic analysis in the case using only the phases when the line profiles of the two components do not overlap.}

\label{table:less_profiles}

\centering
\begin{tabular}{c c c c c c c c c}
\hline\hline
component & $B_{mean}$ & $B_{max}$ & poloidal & toroidal & dipole   & quadrupole & octupole & axisymmetric \\
          &  [G]       &   [G]      & [\% tot] & [\% tot] & [\% pol] & [\% pol]  & [\% pol] & [\% tot]     \\
\hline
\\
primary   & 197.3 & 589.5 & 93.7 & 6.3 & 66.6 & 14.4 & 9.8 & 73.9 \\
secondary & 238.0 & 580.1 & 97.3 & 2.7 & 88.3 & 7.0 & 3.8 & 71.0  \\
\\
\hline
\hline
\end{tabular}
\end{center}
\end{table*}

\vspace{4cm}

\section{ZDI with inflated $v\sin i$ values} 
\label{appendix:ZDI_model_vsini_inflated}

\begin{figure*} [h!]
    \sidecaption
    \includegraphics[width=0.3\textwidth]{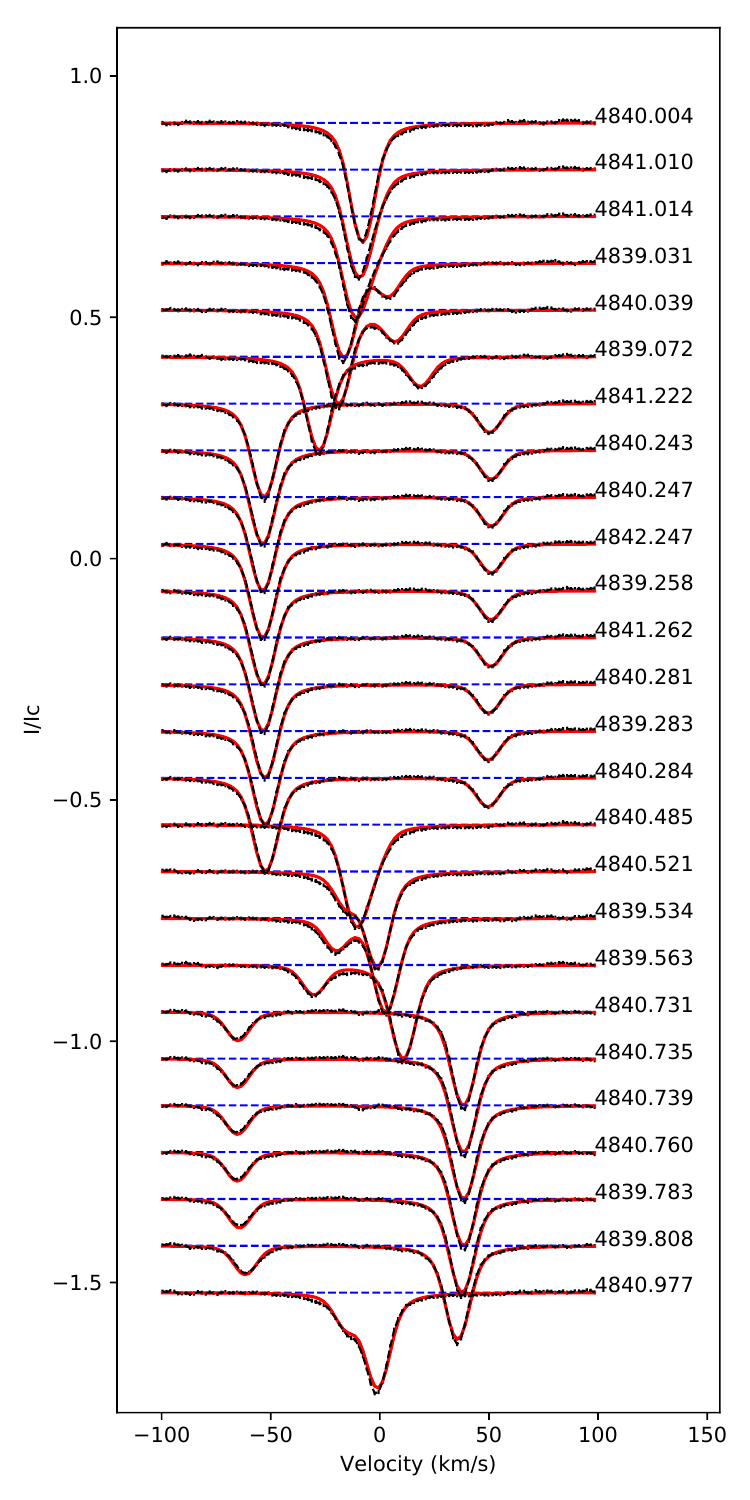}
    \includegraphics[height=0.45\textheight]{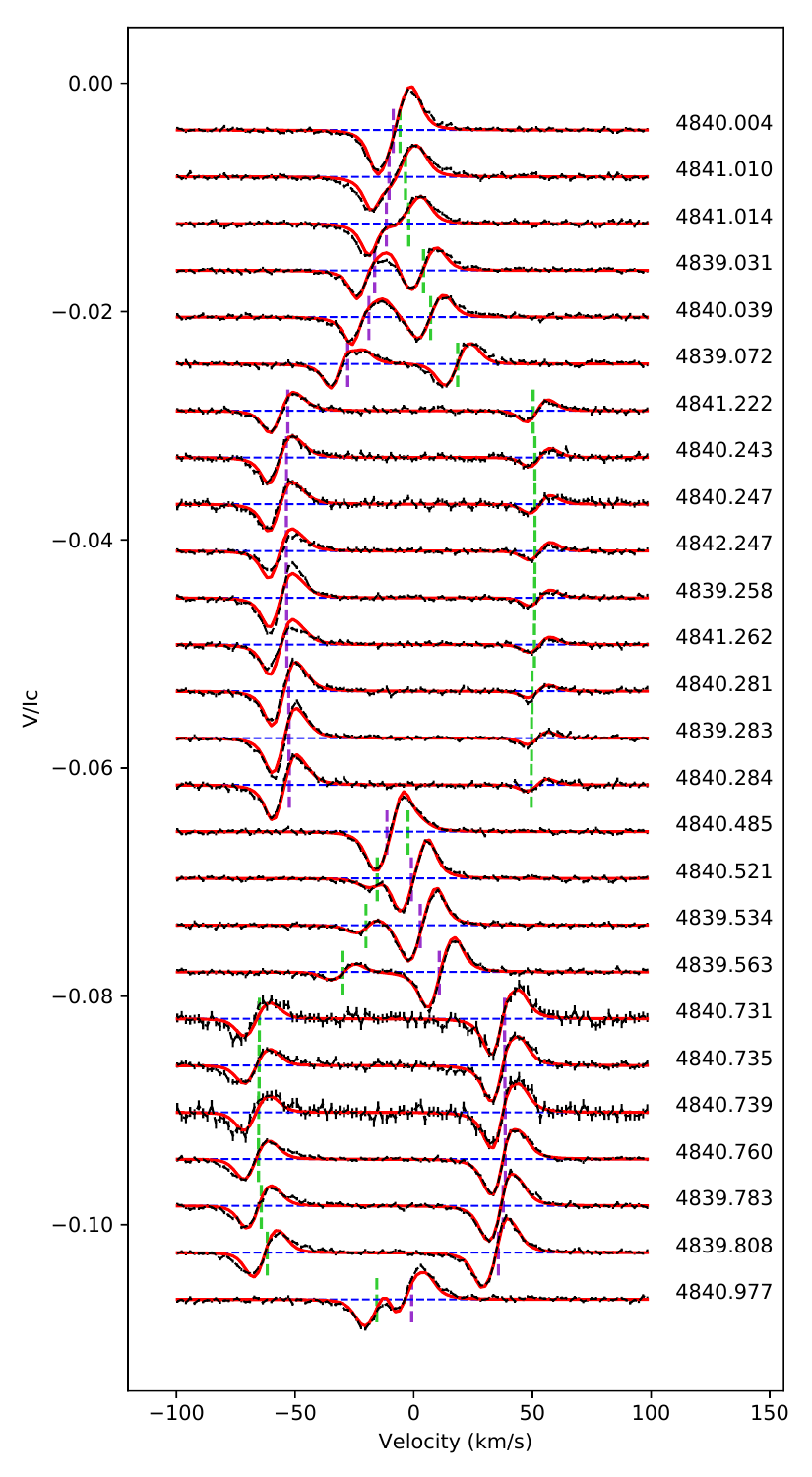}
     \caption{Normalised Stokes $I$ and $V$ profiles of FK~Aqr in the case of inflated radii of both components by 30~\%.}
   \label{fig:stokesIV_inflated_vsini}
   \end{figure*}

\begin{figure*} [h!]
    \centering
    \includegraphics[]{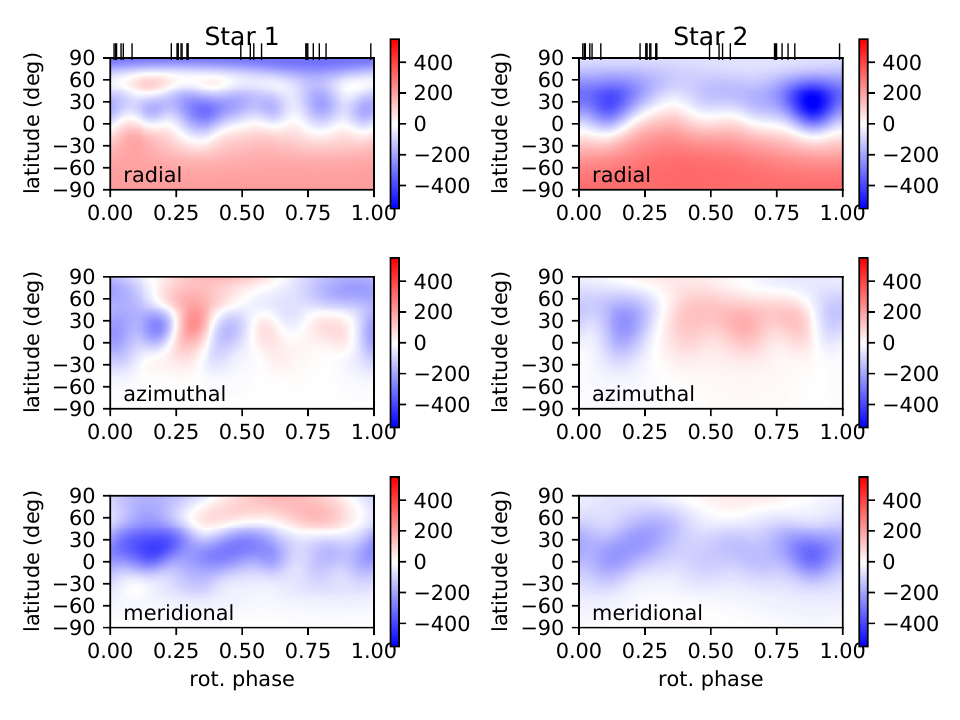}
     \caption{Magnetic maps of the system FK~Aqr in the case of inflated radii of both components by 30~\%.}
   \label{fig:zdi_inflated_vsini}
   \end{figure*}

We explore the case in which the radii of the two components of FK~Aqr are inflated by 30\%. Thus, the $v\sin i$ values for the primary and secondary become 7.05~km\,s$^{-1}$ and 5.72~km\,s$^{-1}$, respectively. We then repeated the procedure described in Appendix~\ref{appendix:ZDI_models} to find the three line parameters for each star. We reconstructed the ZDI models. The magnetic characteristics are given in Table~\ref{table:models_inflated_vsini}. The corresponding fit of Stokes $I$ and $V$ profiles, and the magnetic maps of the two components are given in Fig.~\ref{fig:stokesIV_inflated_vsini} and Fig.~\ref{fig:zdi_inflated_vsini}.


\begin{table*} [h!]

\centering
\begin{center}

\caption{Magnetic analysis in the case where the radii of both components are  inflated by 30\%.}

\label{table:models_inflated_vsini}

\centering
\begin{tabular}{c c c c c c c c c}
\hline\hline
component & $B_{mean}$ & $B_{max}$ & poloidal & toroidal & dipole   & quadrupole & octupole & axisymmetric \\
          &  [G]       &   [G]      & [\% tot] & [\% tot] & [\% pol] & [\% pol]  & [\% pol] & [\% tot]     \\
\hline
\\
primary & 192--199 & 419--480 & 90--91 & 9--10 & 67--75 & 12--13 & 7--9 & 72--75 \\
secondary & 231--255 & 529--589 & 97--98 & 2--3 & 83--89 & 6--9 & 4--6 & 74--76 \\
\\

\hline
\hline
\end{tabular}
\end{center}
\end{table*}

\end{appendix}
\end{document}